\documentclass[twocolumn,epjc3]{svjour3}  
\smartqed  
\RequirePackage{graphicx}
\usepackage{lineno}
\usepackage{xcolor}

\usepackage{amssymb}
\usepackage{amsmath}
\RequirePackage[numbers,sort&compress]{natbib}
\RequirePackage[colorlinks,citecolor=blue,urlcolor=blue,linkcolor=blue]{hyperref}
\journalname{Eur. Phys. J. A}
\newcommand{\Eq}[1]   {Eq.~(\ref{#1})}

\newcommand{\Fi}[1]   {Fig.~\ref{#1}}
\newcommand{\Fis}[2]  {Figs.~\ref{#1} and~\ref{#2}}

\newcommand{\Ta}[1]   {Tab.~\ref{#1}}

\newcommand{\Se}[1]   {Sect.~\ref{#1}}

\newcommand{\agev}    {\mbox{~$A$GeV}}               
\newcommand{\gevc}    {\mbox{GeV$/c$}}

\newcommand{\mevc}    {\mbox{MeV$/c$}}

\newcommand{\mum}     {\mbox{$\mu$m}}

\newcommand{\rb}[1]   {\mbox{\textrm{\scriptsize #1}}}
\newcommand{\rbt}[1]  {\mbox{\textrm{\tiny #1}}}
\newcommand{\hefour}  {\ensuremath{{}^{4}\textrm{He}}}

\newcommand{\sqrtsnn} {\ensuremath{\sqrt{s_{_{\rbt{NN}}}}}}

\newcommand{\pt}      {\ensuremath{p_{\rb{t}}}}

\newcommand{\dedx}    {\ensuremath{\textrm{d}E/\textrm{d}x}}

\newcommand{\bav}     {\ensuremath{\langle b \rangle}}

\newcommand{\ebeam}   {\ensuremath{E_{\rb{beam}}}}
\newcommand{\yproj}   {\ensuremath{y_{\rb{proj}}}}
\newcommand{\ycm}     {\ensuremath{y_{\rb{cm}}}}

\newcommand{\chisq}   {\ensuremath{\chi^{2}}}

\newcommand{\etwo}    {\ensuremath{\epsilon_{2}}}
\newcommand{\etwoav}  {\ensuremath{\langle \epsilon_{2} \rangle}}
\newcommand{\efour}   {\ensuremath{\epsilon_{4}}}
\newcommand{\efourav} {\ensuremath{\langle \epsilon_{4} \rangle}}
\newcommand{\vn}      {\ensuremath{v_{n}}}
\newcommand{\vone}    {\ensuremath{v_{1}}}
\newcommand{\vtwo}    {\ensuremath{v_{2}}}
\newcommand{\vthree}  {\ensuremath{v_{3}}}
\newcommand{\vfour}   {\ensuremath{v_{4}}}
\newcommand{\vfive}   {\ensuremath{v_{5}}}

\newcommand{\dvonedyp}{\ensuremath{d \vone/d y^{\prime}}}
\newcommand{\dvonemid}{\ensuremath{d \vone/d y^{\prime}|_{y^{\prime} = 0}}}

\newcommand{\dvthrdyp}{\ensuremath{d \vthree/d y^{\prime}}}
\newcommand{\dvthrmid}{\ensuremath{d \vthree/d y^{\prime}|_{y^{\prime} = 0}}}
\newcommand{\psirp}   {\ensuremath{\Psi_{\rb{RP}}}}
\newcommand{\psiep}   {\ensuremath{\Psi_{\rb{EP}}}}
\newcommand{\psiepone}{\ensuremath{\Psi_{\rb{EP,1}}}}
\newcommand{\psiepn}  {\ensuremath{\Psi_{\rb{EP,n}}}}

\begin{document}
%

%
\title{Proton, deuteron and triton flow measurements in Au+Au
  collisions at $\sqrtsnn = 2.4$~GeV}
%
%
\author{
  HADES collaboration \\[5bp]
  J.~Adamczewski-Musch$^{5}$,
  O.~Arnold$^{10,9}$,
  C.~Behnke$^{8}$,
  A.~Belounnas$^{13}$,
  J.C.~Berger-Chen$^{10,9}$,
  A.~Blanco$^{2}$,
  C.~Blume$^{8}$,
  M.~B\"{o}hmer$^{10}$,
  P.~Bordalo$^{2}$,
  L.~Chlad$^{14}$
% ...
}

\institute{
  \mbox{Institute of Physics, Slovak Academy of Sciences,
    84228~Bratislava, Slovakia} \\
  \mbox{$^{2}$~LIP-Laborat\'{o}rio de Instrumenta\c{c}\~{a}o e
    F\'{\i}sica Experimental de Part\'{\i}culas, 3004-516~Coimbra, Portugal} \\
  \mbox{$^{3}$~Institute of Nuclear Physics, Polish Academy of
    Sciences, 31342 Krak\'{o}w, Poland} \\
  \mbox{$^{4}$~Smoluchowski Institute of Physics, Jagiellonian University
    of Cracow, 30-059~Krak\'{o}w, Poland} \\
  \mbox{$^{5}$~GSI Helmholtzzentrum f\"{u}r Schwerionenforschung GmbH,
    64291~Darmstadt, Germany} \\
  \mbox{$^{6}$~Technische Universit\"{a}t Darmstadt, 64289~Darmstadt,
    Germany} \\
  \mbox{$^{7}$~Institut f\"{u}r Strahlenphysik, Helmholtz-Zentrum
    Dresden-Rossendorf, 01314~Dresden, Germany} \\
  \mbox{$^{8}$~Institut f\"{u}r Kernphysik, Goethe-Universit\"{a}t,
    60438 ~Frankfurt, Germany} \\
  \mbox{$^{9}$~Excellence Cluster 'Origin and Structure of the
    Universe', 85748~Garching, Germany} \\
  \mbox{$^{10}$~Physik Department E62, Technische Universit\"{a}t
    M\"{u}nchen, 85748~Garching, Germany} \\
  \mbox{$^{11}$~II.Physikalisches Institut, Justus Liebig
    Universit\"{a}t Giessen, 35392~Giessen, Germany} \\
  \mbox{$^{12}$~Department of Physics, University of Cyprus,
    1678~Nicosia, Cyprus} \\
  \mbox{$^{13}$~Laboratoire de Physique des 2 infinis Ir\`{e}ne
    Joliot-Curie, Universit\'{e} Paris-Saclay, CNRS-IN2P3., F-91405 Orsay, France} \\
  \mbox{$^{14}$~Nuclear Physics Institute, The Czech Academy of
    Sciences, 25068~Rez, Czech Republic} \\
  \mbox{$^{15}$~LabCAF. F. F\'{\i}sica, Univ. de Santiago de
    Compostela, 15706~Santiago de Compostela, Spain} \\
  \mbox{$^{16}$~Uniwersytet Warszawski, Wydzia\l\ Fizyki, Instytut
    Fizyki Do\'{s}wiadczalnej, 02-093~Warszawa, Poland} \\
  \mbox{$^{17}$~Dipartimento di Fisica and INFN, Universit\`{a}
    di Torino, 10125~Torino, Italy} \\
  \mbox{$^{18}$~Frederick University, 1036~Nicosia, Cyprus} \\
  \\
  \\
  \mbox{$^{a}$ also at Coimbra Polytechnic - ISEC, ~Coimbra, Portugal}
  \\
  \mbox{$^{b}$ also at Technische Universit\"{a}t Dresden,
    01062~Dresden, Germany} \\
  \mbox{$^{\dagger}$ deceased} \\
} 
%
\date{Version 3.3: \today}
%
\maketitle
%

%
\begin{abstract}
High-precision measurements of flow coefficients \vn\ ($n = 1 - 4$) for
protons, deuterons and tritons relative to the first-order spectator
plane have been performed in Au+Au collisions at $\sqrtsnn = 2.4$~GeV
with the High-Acceptance Di-Electron Spectrometer \linebreak (HADES)
at the SIS18/GSI. Flow coefficients are studied as a function of
transverse momentum \pt\ and rapidity \ycm\ over a large region of
phase-space and for several classes of collision centrality. A clear
mass hierarchy, as expected by relativistic hydrodynamics, is found
for the slope of \vone, $\dvonedyp|_{y^{\prime} = 0}$ where
$y^{\prime}$ is the scaled rapidity, and for \vtwo\ at mid-rapidity.
Scaling with the number of nucleons is observed for the \pt~dependence
of \vtwo\ and \vfour\ at mid-rapidity, which is indicative for nuclear
coalescence as the main process responsible for light nuclei
formation.  \vtwo\ is found to scale with the initial eccentricity
\etwoav, while \vfour\ scales with $\etwoav^{2}$ and \efourav.  The
multi-differential high-precision data on \vone, \vtwo, \vthree, and
\vfour\ provides important constraints on the equation-of-state of
compressed baryonic matter.
\end{abstract}
%

%
\section{Introduction}
\label{sect:intro}

Heavy-ion collisions are a tool to investigate the properties of
strongly-interacting matter under extreme conditions, such as high
temperatures typical for the early phase of the universe and high
baryon number densities occurring in compact stellar objects
\cite{Adamczewski-Musch:2019byl}.  Especially the latter conditions
cannot easily be addressed by ab-initio calculations based on Quantum
Chromo-Dynamics (QCD), the theory of strong interaction, but are to be
addressed with effective theories.  Therefore, measurements are
indispensable to determine the properties of dense matter, which can
be -- in local equilibrium -- encoded in the equation-of-state (EOS)
\cite{Danielewicz:2002pu,Fevre:2015fza,Huth:2021bsp}. 

For non-polarised and equal-size projectile and target nuclei, the
azimuthal uniformity of the momentum tensor in the final state is
broken by the spatial asymmetry of the initial state at finite impact
parameter which is transferred into momentum space via pressure
gradients generated by multiple interactions of the matter
constituents.  The resulting structure of the azimuthal distributions
of produced particles is conveniently parametrised by a Fourier
decomposition \cite{Voloshin:1994mz} with the coefficients
$\vn(\pt,y)$:
\begin{equation}
\label{eq:fourier}
E \frac{d^{3} N}{d p^{3}} = \frac{d^{2} N}{\pi dy d(\pt^{2})}
                           [1 + 2 \sum_{n = 1}^{\infty} \vn(\pt,y) 
                            \cos n (\phi - \psirp)] \, . 
\end{equation}
Here, $\phi$ is the azimuthal angle of the particle and \psirp\ stands
for the azimuthal angle of the reaction plane, defined by the beam
direction $\vec{z}$ and the direction of the impact parameter $\vec{b}$
of the colliding nuclei. $p_{z} = p \, \cos \theta$ is the momentum
component along the beam direction $\vec{z}$ with the laboratory
momentum $p$ and the polar angle $\theta$, $\pt = p \, \sin \theta$ the
one perpendicular to it, and $y = \tanh^{-1} (p_{z} / E)$ denotes the
rapidity of a given particle with energy $E$ in the laboratory frame.
The rapidity in the centre-of-mass system is denoted by $\ycm = y - \,
\yproj / 2$, with the projectile rapidity \yproj. The coefficients
\vone\ and \vtwo\ quantify the so-called directed and elliptic flow,
respectively. More generally, the shape of the anisotropy is quantified
by odd coefficients \vone, \vthree, \vfive\ $\dots$ $v_{2n+1}$ and even
coefficients \vtwo, \vfour\ $\dots$ $v_{2n}$. Due to momentum
conservation, and assuming that initial-state fluctuations with large
eccentricities are absent, the values of even coefficients are expected
to be symmetric around mid-rapidity, while the ones of the odd
harmonics should change their sign when going from forward to backward
centre-of-mass rapidities.

%
\begin{figure}
\begin{center}
\includegraphics[width=1.0\linewidth]{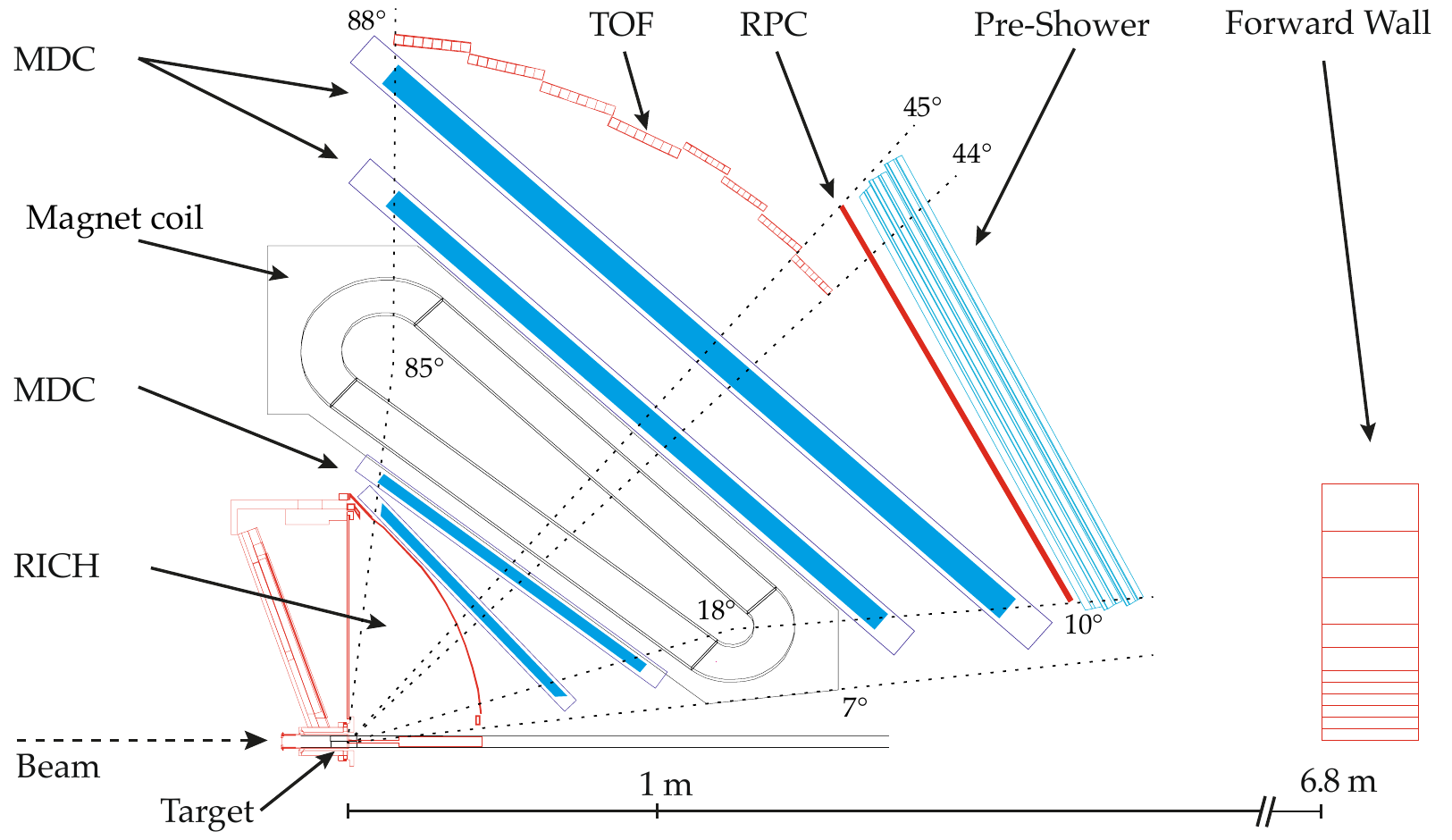}
\end{center}
\caption{Cross section of the HADES set-up during the measurement of
  Au+Au collisions at $\sqrtsnn = 2.4$~GeV.  Shown is the arrangement
  of the different detectors and a magnet coil on one side of the beam
  pipe.}
\label{fig:hades}
\end{figure}
%

%
\begin{figure*}
\begin{center}
\includegraphics[width=1.0\textwidth]{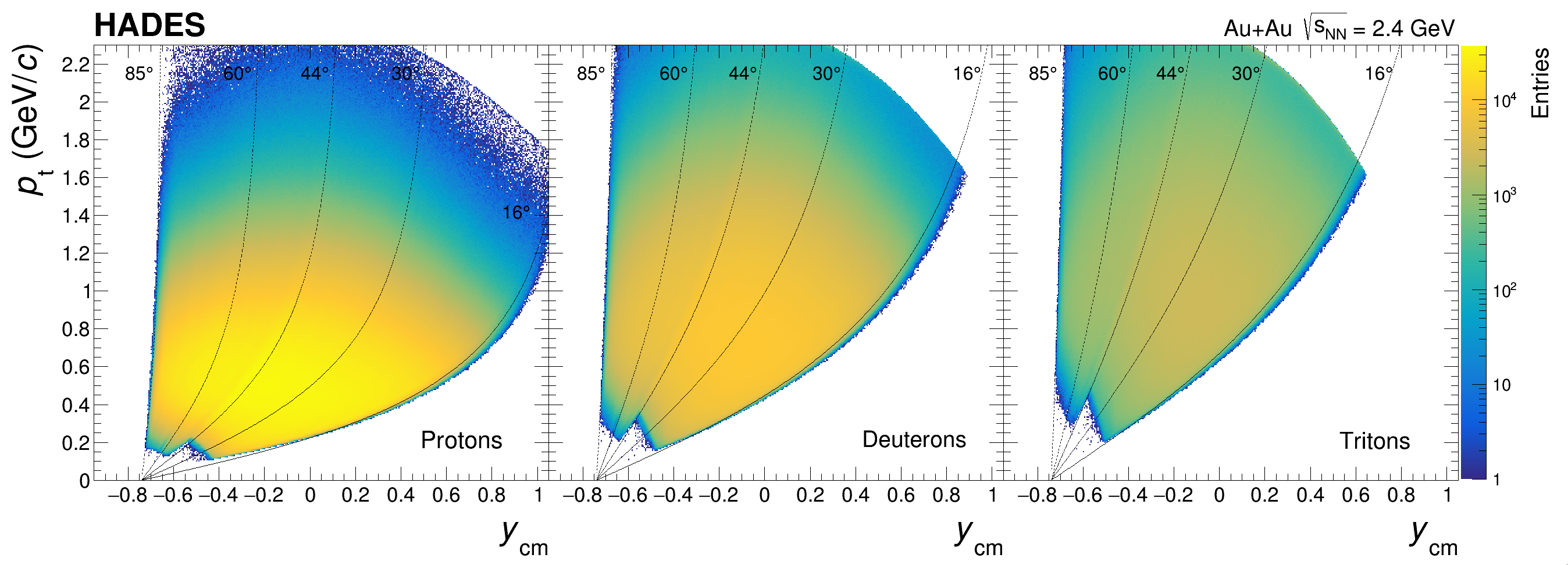}
\end{center}
\caption{The phase-space coverage of identified protons (left panel),
  deuterons (middle panel) and tritons (right panel) as accepted in the
  HADES experiment for Au+Au collisions at $\sqrtsnn = 2.4$~GeV as a
  function of the centre-of-mass rapidity \ycm\ and transverse momentum
  \pt\ (the curves correspond to the given polar angles $\theta$ in the
  laboratory system).}
\label{fig:PtvsY-p-d-t}
\end{figure*}
%

Flow data in the few GeV energy range at BEVALAC and SIS18 have
been reported for pions, charged kaons, hyperons, neutrons, as well as
protons and many light nuclei.  For reviews see 
\cite{Ritter:2014uca,Andronic:2006ra,Herrmann:1999wu,Reisdorf:1997fx} 
and references therein.  Most of these data are for a limited phase
space or integrated over transverse momenta.  High-statistics,
multi-differential data on \vone\ and \vtwo\ for identified particles
measured over a large region of phase-space is a valuable extension of
the existing world data.  In addition, the study of higher-order flow
coefficients can provide information about the various contributions
to the bulk properties of dense nuclear matter.  At RHIC and LHC
energies, these were employed to determine the ratios of the shear and
bulk viscosity to the entropy density $\eta/s$, and $\zeta/s$,
respectively, of high-temperature matter
\cite{Heinz:2013th,Ryu:2015vwa}.  Attempts to extract $\eta/s$
for dense baryonic matter have been made by comparing various model
approaches to the available data
\cite{Demir:2008tr,Khvorostukhin:2010aj,Barker:2016hqv,Ivanov:2016hes,Rose:2017bjz,Reichert:2020oes},
however with large uncertainties.  Data on higher-order flow
coefficients are essential in order to disentangle the effects of
shear and bulk viscosity \cite{Reichert:2020oes} and can also provide
important information on the EOS.  A comparison of the proton \vthree,
measured by HADES, with UrQMD transport model calculations indicates
an enhanced sensitivity to the EOS of the hadronic medium
\cite{Hillmann:2018nmd,Hillmann:2019wlt}. Other transport model
calculations suggest that a non-vanishing value of \vfour, measured at
center-of-mass energies of a few GeV, can constrain the nuclear mean
field at high net-baryon densities \cite{Danielewicz:1999zn}. The E877
collaboration has reported a non-vanishing \vfour\ at 10.1\agev\
\cite{Barrette:1994xr}, but measurements of higher coefficients are
generally scarce at low (i.e. AGS and SIS18) energies. A
multi-differential measurement of several Fourier coefficients allows
for a three-dimensional characterization of heavy-ion collisions in
different representations \cite{Adamczewski-Musch:2020iio,
Danielewicz:2021vqq, Reichert:2022yxq}.

The scaling properties of the flow coefficients \vn\ of different
order $n$ with the number of nucleons $A$ of the respective nucleus
can provide information on the production mechanisms of light nuclei,
e.g. via nucleon coalescence \cite{Molnar:2003ff, Kolb:2004gi}.  The 
relation of the \vn\ to the shape of the initial eccentricity of the
collision system can shed light on the reaction dynamics and the
transport properties of the produced medium.

In this article we present results on flow of protons, deuterons 
and tritons in Au+Au collisions at a centre-of-mass energy of
$\sqrtsnn = 2.4$~GeV, corresponding to a kinetic beam energy of
1.23\agev.  We extend our previous study
\cite{Adamczewski-Musch:2020iio} to multi-differential data on the
flow coefficients $\vone - \vfour$ as a function of transverse
momentum and rapidity over a large region of phase-space and for
several classes of reaction centrality.

The paper is organised as follows. Section~\ref{sect:experiment}
introduces the experimental set-up and the particle reconstruction
methods, while section~\ref{sect:flow} discusses the procedures used
to determine the flow coefficients.  Section~\ref{sect:vn} presents
the results on directed, elliptic and higher harmonics for
semi-central collisions.  In section~\ref{sect:comparison} our results
of \vone\ and \vtwo\ are compared with existing world data, while in
section~\ref{sect:scaling} the scaling properties of the data are
discussed.  Section~\ref{sect:models} presents comparisons to several
model calculations.  We summarize in section~\ref{sect:conclusions}.
%

%
\section{Experimental set-up}
\label{sect:experiment}

HADES is a charged-particle detector consisting of a six-coil toroidal
superconducting magnet centered around the beam axis with six identical
detection sections located between the coils and covering polar angles
between $18^{\circ}$ and $85^{\circ}$ (see \Fi{fig:hades}).  Each sector
is equipped with a Ring-Imaging Cherenkov (RICH) detector followed by
four layers of Multi-Wire Drift Chambers \linebreak (MDCs), two in
front of and two behind the magnetic field, as well as a scintillator
Time-Of-Flight detector (TOF) (polar angle coverage:
$44^{\circ}$~--~$85^{\circ}$) and Resistive Plate Chambers (RPC)
(polar angle coverage: $18^{\circ}$~--~$45^{\circ}$). Hadron
identification is based on the time-of-flight measured with TOF and
RPC, and on the energy-loss information from TOF as well as from the
MDC tracking chambers.  Electron candidates are in addition selected
via their signals in the RICH detector.  Combining this information
with the track momentum, as determined from the deflection of the
tracks in the magnetic field, allows for an identification of charged
particles (e.g. pions, kaons or protons).

The spectrometer set-up is complemented by the Forward Wall (FW)
detector. It consists of 288 scintillator elements of $2.54$~cm
thickness and three different front area sizes (innermost region: $4
\times 4$~cm$^{2}$, intermediate region: $8 \times 8$~cm$^{2}$ and
outermost region: $16 \times 16$~cm$^{2}$) which are read out with
photomultipliers. The FW is placed downstream at a $6.8$~m distance
from the target and covers the polar angles $0.34^{\circ} < \theta <
7.4^{\circ}$.  It thus allows for a measurement of the emission angles
and the charge states of projectile spectators and is used to
determine the event-plane angle. A detailed description of the HADES
experiment can be found in \cite{Agakishiev:2009am}.

\subsection{Data set and event selection}
\label{sect:event_selection}

Several triggers are implemented to start the data acquisition.  The
minimum-bias trigger is defined by a signal in a 60~\mum\ thick 
mono-crystalline CVD\footnote{Chemical Vapor Deposition} diamond 
detector (START) in the beam line \cite{Pietraszko:2014tba}.  In 
addition, online Physics Triggers (PT) are used, which are based 
on hardware thresholds on the TOF signals, proportional to the 
event multiplicity, corresponding to at least 5 (PT2) or 20 (PT3)
hits in the TOF. 

By comparing the measured hit multiplicity distribution with Glauber
and transport model simulations, it has been estimated that the
PT3~trigger is selecting $(43 \pm 4)$~\% (PT2~trigger: $(72 \pm
4)$~\%) of the total inelastic cross section
\cite{Adamczewski-Musch:2017sdk}.  The selection of centrality classes
is based on the summed hit multiplicity of the TOF and RPC detectors.
Four classes are defined, which together cover the $40$~\% most
central collisions in steps of $10$~\% of the total Au+Au cross
section of $6.83 \pm 0.43$~b.  Events are selected offline by
requiring that their reconstructed global event vertex lies inside the
target region, i.e. between $z = -65$~mm and $0$~mm along the beam
axis.  Additionally, only events with at least four hits in the FW
with a charge $Z \ge 1$ are used for the reconstruction of the
event-plane.  It was verified that this criterion does not introduce
any significant bias to the centrality selection.  The mean number of
accepted proton, deuteron and triton candidates are summarized in
\Ta{tab:evt_stat}.

%
\begin{table}
\begin{tabular*}{\linewidth}{@{\extracolsep{\fill}}rrrr@{}}
\hline
Centrality                                      &
\multicolumn{1}{c}{$\langle M_{\rb{p}} \rangle$} &
\multicolumn{1}{c}{$\langle M_{\rb{d}} \rangle$} &
\multicolumn{1}{c}{$\langle M_{\rb{t}} \rangle$} \\
\hline
 $ 0 - 10$~\% & 27.8 &  9.9 &  2.4 \\  
 $10 - 20$~\% & 19.7 &  7.0 &  1.9 \\  
 $20 - 30$~\% & 13.7 &  4.7 &  1.3 \\
 $30 - 40$~\% & 10.5 &  3.4 &  0.9 \\
\hline
\end{tabular*}
\caption{The mean multiplicities of accepted proton ($\langle
  M_{\rb{p}} \rangle$), deuteron ($\langle M_{\rb{d}} \rangle$) and
  triton ($\langle M_{\rb{t}} \rangle$) candidates (uncorrected raw
  numbers) for the different centrality classes.}
\label{tab:evt_stat}
\end{table}
%

\subsection{Track selection}
\label{sect:track_selection}

The flow coefficients are determined for the charged particles
detected by the detectors MDC, TOF and RPC.  Their trajectories are
reconstructed using the MDC information.  The resulting tracks are
selected according to the quality parameter provided by the
employed Runge-Kutta track fitting algorithm $\chi^{2}_{\rb{RK}}$ 
and a maximal Distance of Closest Approach ($DCA$) of the extrapolated
track to the reconstructed primary vertex position.  In order to
assure a good matching of the tracks to the hits measured in the
particle identification detectors TOF and RPC, an additional selection
criterion is applied.  It involves an upper limit on the quality
parameter $\rb{Q}_{\rb{MM}} = dx / \sigma_{x}$, which is defined as the
deviation of the intersection point of a given reconstructed track
from the position of the associated hit in the RPC and TOF detectors,
$dx$, normalized to the corresponding measurement uncertainty,
$\sigma_{x}$.  The nominal selection values on $\chi^{2}_{\rb{RK}}$, 
$DCA$ and $\rb{Q}_{\rb{MM}}$ are summarized in \Ta{tab:cut-parameter}.

%
\begin{table}[t]
\begin{tabular*}{\linewidth}{@{\extracolsep{\fill}}llll@{}}
\hline
Selection criterion          &
Nominal                      &
\multicolumn{2}{c}{Variations}\\

\hline
$\chi^{2}_{\rb{RK}}$      & $ < 1000$     & $ < 200$   & $ < 15$      \\
$\rb{Q}_{\rb{MM}}$        & $ < 3$        & $ < 0.5$   & $ > 0.5$     \\
$DCA$                     & $ < 10$~mm    & $ < 8$~mm  & $ > 2$~mm    \\
\hline
$n \, \sigma_{\beta}(p)$  &  2.5          & 3.5        &  4.5         \\
$Z_{\rb{MDC}}$ protons    & -0.25 - 0.75  & \multicolumn{2}{c}{-0.50 - 1.00} \\
$Z_{\rb{MDC}}$ deuterons  & -0.25 - 0.50  & \multicolumn{2}{c}{-0.50 - 0.75} \\
$Z_{\rb{MDC}}$ tritons    & -0.25 - 0.50  & \multicolumn{2}{c}{-0.50 - 0.75} \\
\hline
\end{tabular*}
\caption{List of applied track selection and PID criteria.  In addition to the
  nominally applied values, also two variations are given for each
  selection criterion, which are used for the determination of the
  systematic uncertainties.}
\label{tab:cut-parameter}
\end{table}
%

\subsection{Particle identification}
\label{sect:pid}

The Particle IDentification (PID) is based on a combined measurement
of time-of-flight and energy loss.  The time-of-flight, as determined
by the TOF and RPC detectors, allows for a separation of particles in
different momentum dependent regions of velocity $\beta$.  To select
protons, deuterons and tritons windows with widths of
$n\:\sigma_{\beta}(p)$ with $n = 2.5$ are placed around the
corresponding expected $\beta$~values (see also
\Ta{tab:cut-parameter}).  The respective resolutions
$\sigma_{\beta}(p)$ depend on the particle momenta $p$ and are
para\-me\-tri\-sed accordingly.

In addition, the energy loss (\dedx) measurements in the MDCs are
employed for PID.  This is particularly important to suppress the
\hefour~contamination in the deuteron sample, as the two nuclei cannot
be separated by time-of-flight alone due to the same $Z/A$~ratio.  The
variable $Z_{\rb{MDC}}$ is constructed from the energy loss measured
in all four MDC layers, $\dedx_{\rb{exp}}$, and the theoretically
expected value, $\dedx_{\rb{th}}$, 
\begin{equation}
  Z_{\rb{MDC}} = \ln \frac{\dedx_{\rb{exp}}}{\dedx_{\rb{th}}}\, .
\end{equation}
The selection windows applied to this variable are momentum 
independent, but different for protons, deu\-te\-rons and tritons, 
see \Ta{tab:cut-parameter}.

The purity of the particle identification procedure is determined by
analysing simulated data (see \Se{sect:efficiency}) and by fitting the
mass distributions, calculated from the measured values of $\beta$ and
momentum for different rapidity and transverse momentum intervals,
with a function that describes the signal as well as the background
component.  Phase-space intervals, in which the purity of the particle
identification is found to be lower than 80~\%, are excluded from
further analysis.  Also, intervals at the edges of the detector
acceptance, i.e. on the borders of the polar angle range $16^{\circ} <
\theta < 85^{\circ}$ and the gaps in azimuth between the detector
sectors, are excluded.  This translates into a rapidity dependent
lower transverse momentum cut-off.  At very high momenta, phase-space
regions are rejected if the accuracy of the momentum measurement is
not sufficient.  The phase-space coverage for the identified particles
is shown in \Fi{fig:PtvsY-p-d-t}.
%

%
\section{Determination of flow coefficients}
\label{sect:flow}

The flow coefficients \vn\ of order $n$ are defined in their relation
to the reaction plane angle \psirp\ as
\cite{Ollitrault:1997di,Poskanzer:1998yz}
\begin{equation}
  \vn = \langle \cos[n (\phi - \psirp)] \rangle \, .
\end{equation}
Here, $\langle \dots \rangle$ denotes the average over all selected
particles and events in a given sample.  As the reaction plane is not
accessible experimentally, it is replaced by the event-plane angle
(\psiep) constructed from measured event anisotropies as described in
the following.

\subsection{Reaction plane reconstruction}

%
\begin{figure}
\begin{center}
\includegraphics[width=1.0\linewidth]{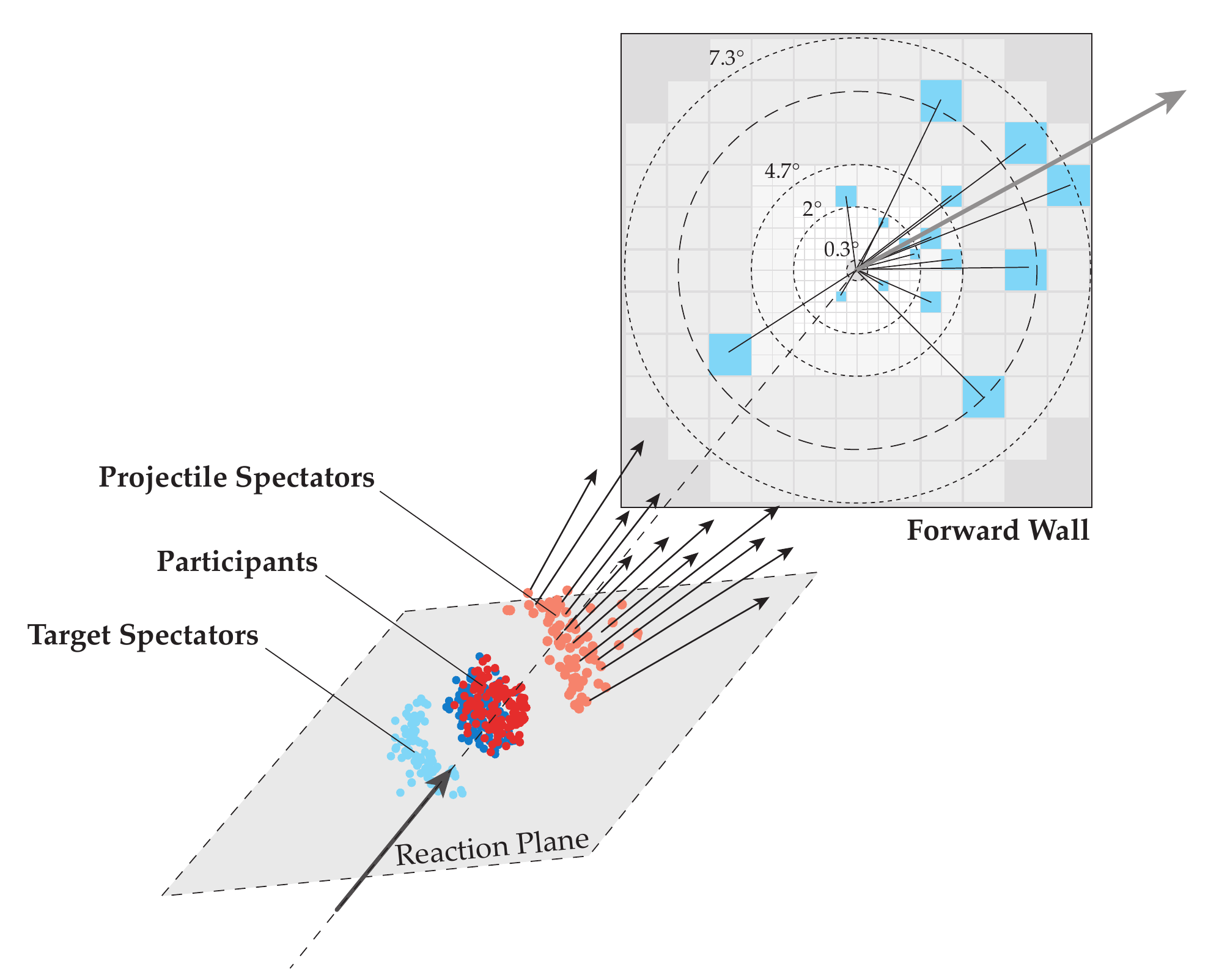}
\end{center}
\caption{Sketch illustrating the event-plane reconstruction using
  the projectile spectator hits recorded in the Forward Wall.}
\label{fig:fw_plane}
\end{figure}
%

For the determination of the event-plane angle \psiep, hits of
projectile spectators recorded by the FW are used as illustrated in
\Fi{fig:fw_plane}.  Projectile spectators are thus measured in the
polar angle interval $0.34^{\circ} < \theta < 7.4^{\circ}$.  Only
those hits are used for which the energy deposit in the scintillator
cells and their flight time corresponds to the values expected for
spectators with charges of $Z \ge 1$.

%
\begin{figure}
\begin{center}
\includegraphics[width=1.0\linewidth]{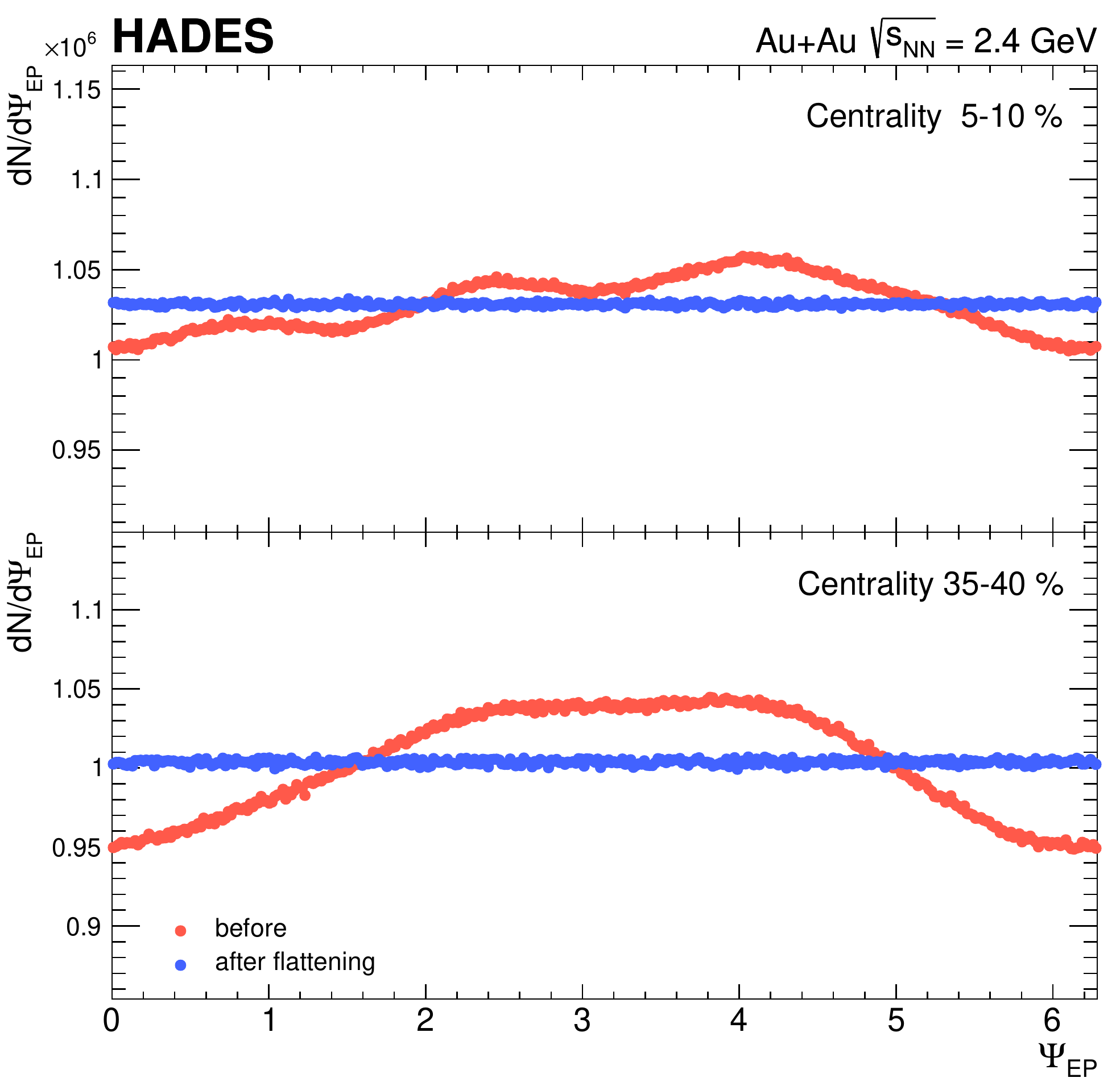}
\end{center}
\caption{The distribution of the first-order event-plane angles 
  \psiepone\ for central ($5 - 10$~\%, upper panel) and 
  semi-central ($35 - 40$~\%, lower panel) Au+Au collisions at 
  $\sqrtsnn = 2.4$~GeV.  Shown are the distributions before (red
  circles) and after (blue circles) applying the flattening procedures
  described in the text.}
\label{fig:EPdistribution}
\end{figure}
%

From the azimuthal angles $\phi_{\rb{FW}}$ of the FW cells hit by
spectators, a vector $\vec{Q}_{n} = (Q_{n,x}, Q_{n,y})$ is calculated
event-by-event:
\begin{eqnarray}
  Q_{n,x} & = & \sum^{N_{\rb{FW}}}_{i = 1}
  w_{i} \; \cos (n \, \phi_{\rb{FW,i}}) \, ,
  \nonumber \\
  Q_{n,y} & = & \sum^{N_{\rb{FW}}}_{i = 1} 
  w_{i} \; \sin (n \, \phi_{\rb{FW,i}}) \, .
\end{eqnarray}
Here, $N_{\rb{FW}}$ is the number of detected FW cell hits.  The
weights $w_{i}$ are here chosen to be $w_{i} = |Z_{i}|$, where $Z_{i}$
is the charge of a given hit as determined via the signal amplitude
seen by the FW cell. Because of non-uniformities in the FW acceptance,
caused by few dead cells and by deviations of the beam position
relative to the nominal centre of the experimental set-up, the
distribution of FW hits averaged over many events is not centred
around the origin. To correct for this, the individual positions of the
FW-hit $X_{\rb{FW},i}$ and $Y_{\rb{FW},i}$ are re-centred by the
corresponding first moments \linebreak $(\langle X_{\rb{FW}} \rangle,
\langle Y_{\rb{FW}} \rangle)$ and scaled by the second moments \linebreak
$(\sigma_{X_{\rb{FW}}}, \sigma_{Y_{\rb{FW}}})$, which are calculated
for each day of data-taking and centrality class separately. To remove
the residual non-uniformities in the event-plane angular distribution
an additional flattening procedure was applied \cite{Barrette:1997pt}.

The corresponding event-plane angle of order $n$ is then defined as:
\begin{equation}
\psiepn = \frac{1}{n} \arctan \frac{Q_{n,y}}{Q_{n,x}} \, .
\end{equation}
Figure~\ref{fig:EPdistribution} shows distributions of the first-order
event-plane angles before and after applying the above described
correction procedure.  As a result of the corrections, \psiepone\ is
distributed uniformly in all centrality classes.  The comparison to a
flat distribution results in $\chisq/NDF$ values in the range $0.83 -
1.09$ for the centralities $0 - 40$~\%.

Generally, \psiepn\ can be determined for each order $n$.  As the
reaction plane orientation is mainly connected to the deflection of
the projectile spectators, $n = 1$ provides the highest resolution
and therefore the first-order event-plane angle \psiepone\ is used 
in the following for the extraction of the observable flow
coefficients of all orders:
\begin{equation}
  \vn^{\rb{obs}} = \langle \cos[n (\phi - \psiepone)] \rangle \, .
\end{equation}

%
\begin{figure}
\begin{center}
\includegraphics[width=1.0\linewidth]{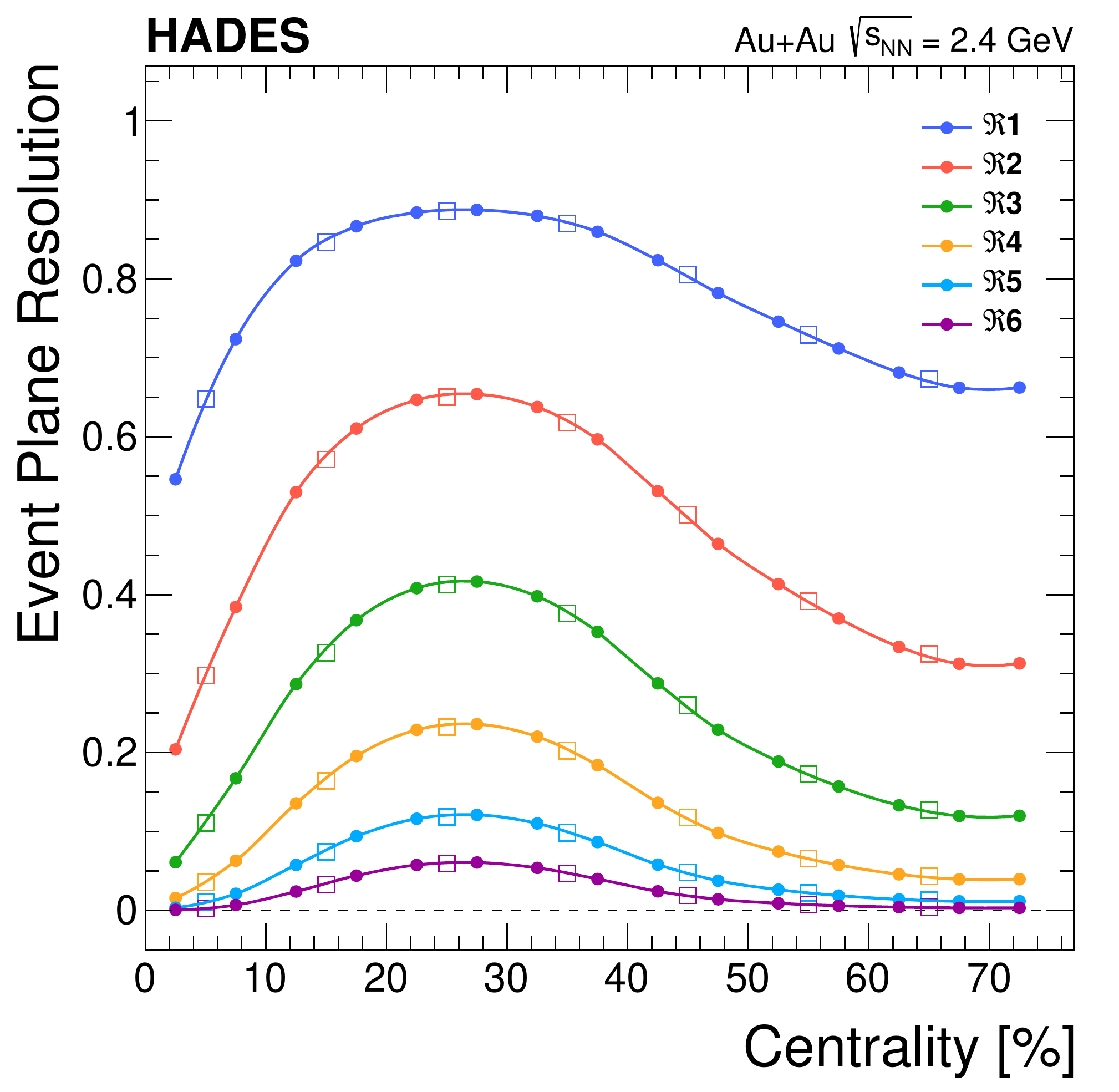}
\end{center}
\caption{The resolution $\Re_{n}$ of the first-order spectator event 
  plane for flow coefficients of different orders $n$ as a function of
  the event centrality \cite{Adamczewski-Musch:2020iio}. The circles
  correspond to centrality intervals of $5$~\% width and the squares to
  $10$~\% width (curves are meant to guide the eye).}
\label{fig:EPresolution}
\end{figure}
%

As the computed event-plane angles will fluctuate around the true
reaction plane angles, the observed flow coefficients will come out
smaller than the true ones.  This can be corrected using the
resolution correction of the event-plane (\psiepone):
\begin{equation}
  \vn = \frac{\vn^{\rb{obs}}}{\Re_{n}} \, .
\end{equation}
For the first-order event-plane, assuming that non-flow contributions 
can be neglected, the resolution can be expressed as 
\cite{Ollitrault:1997di, Ollitrault:1993ba, Poskanzer:1998yz, Borghini:2002mv}
\begin{eqnarray}
\label{eq:voloshin}
\Re_{n} & = & \langle \cos [ n \, (\psiepone - \psirp) ] \rangle 
\nonumber \\
       & = & \frac{\sqrt{\pi}}{2} \, \chi \, e^{- \chi^{2}/2} 
         \left[ I_{\frac{n - 1}{2}} \left(\frac{\chi^{2}}{2}\right) +
                I_{\frac{n + 1}{2}} \left(\frac{\chi^{2}}{2}\right)
         \right] \, ,
\end{eqnarray}
where $I_{\nu}$ are the modified Bessel functions of the order $\nu$
and $\chi$ is the resolution parameter.  For the determination of
$\Re_{n}$, the two-sub-event method is employed.  For this purpose the
FW hits in a given event are randomly divided into two sub-events A
and B of equal multiplicity.  From the correlation of the two the
resolution for the sub-events is then calculated as
\begin{eqnarray}
\Re_{n}^{\rb{sub}} 
  & = & \langle \cos [ n (\Psi_{\rb{EP,A(B)}} - \psirp) ] \rangle
  \nonumber \\
  & = & \sqrt{\langle \cos [ n (\Psi_{\rb{EP,A}} 
                              - \Psi_{\rb{EP,B}}) ] \rangle}
  \;.
\end{eqnarray}
By replacing $\chi$ in \Eq{eq:voloshin} with $\chi^{\rb{sub}}$ and
inverting the equation, the value for $\chi^{\rb{sub}}$ can be
calculated. The value of the resolution parameter for the full FW is
then $\chi = \sqrt{2} \; \chi^{\rb{sub}}$, which yields the full
resolution $\Re_{n}$ after inserting it into \Eq{eq:voloshin}.

The resulting values for the resolution of different order $n$ are
exhibited in \Fi{fig:EPresolution}.  In the case $n = 1$, it is found
to be around $80$~\% and higher in the centrality range $10 - 40$~\%,
while it drops towards a value of $\sim 50$~\% for very central 
collisions.

Alternatively, two methods proposed in \cite{Ollitrault:1997di} are
used. The resolution parameter $\chi$ is obtained by fitting Eq.~(12) of
\cite{Ollitrault:1997di} to the measured distribution of the
differences between the two sub-event-plane angles $\Delta \Psi =
|\Psi_{\rb{EP,A}} - \Psi_{\rb{RP,B}}|$ or by using the approximate
relation
\begin{equation}
\frac{N(\pi/2 < \Delta \Psi < \pi)}{N(0 < \Delta \Psi < \pi)} =
\frac{\exp(- \chi^{2}/2)}{2} \;,
\end{equation}
where the effect of these differences on the systematic
uncertainties is found to be negligible.

%
\begin{table*}[t]
\scriptsize
\begin{tabular*}{\linewidth}{@{\extracolsep{\fill}}r|cc|cc|cc@{}}
 & \multicolumn{2}{c|}{Protons}   
 & \multicolumn{2}{c|}{Deuterons}     
 & \multicolumn{2}{c}{Tritons}   \\
\hline
 & \pt~Dep. & \ycm~Dep. & \pt~Dep. & \ycm~Dep. & \pt~Dep. & \ycm~Dep. \\
\hline
\multicolumn{7}{c}{\vone}\\
\hline
Total syst. uncert. & $0.011 - 0.026$ & $0.012 - 0.022$ & $0.012 - 0.024$ & $0.013 - 0.018$ & $0.016 - 0.027$ & $0.004 - 0.072$\\
\hline
                PID & $0.006 - 0.018$ & $0.008 - 0.018$ & $0.005 - 0.014$ & $0.003 - 0.011$ & $0.013 - 0.024$ & $0.003 - 0.060$\\
      Track Quality & $0.004 - 0.018$ & $0.006 - 0.011$ & $0.003 - 0.014$ & $0.004 - 0.013$ & $0.011 - 0.019$ & $0.004 - 0.021$\\
          Occupancy & $0.013 - 0.021$ & $0.011 - 0.022$ & $0.010 - 0.026$ & $0.007 - 0.019$ & $0.006 - 0.029$ & $0.005 - 0.028$\\
         Acceptance & $0.006 - 0.029$ & $0.013 - 0.024$ & $0.014 - 0.022$ & $0.013 - 0.018$ & $0.017 - 0.031$ & $0.002 - 0.070$\\
\hline
\multicolumn{7}{c}{\vtwo}\\
\hline
Total syst. uncert. & $0.003 - 0.012$ & $0.005 - 0.013$ & $0.004 - 0.017$ & $0.005 - 0.016$ & $0.007 - 0.011$ & $0.006 - 0.027$\\
\hline
                PID & $0.001 - 0.009$ & $0.002 - 0.010$ & $0.001 - 0.011$ & $0.002 - 0.008$ & $0.003 - 0.007$ & $0.004 - 0.024$\\
      Track Quality & $0.002 - 0.013$ & $0.002 - 0.006$ & $0.002 - 0.019$ & $0.002 - 0.009$ & $0.004 - 0.009$ & $0.003 - 0.012$\\
          Occupancy & $0.005 - 0.009$ & $0.005 - 0.016$ & $0.006 - 0.009$ & $0.006 - 0.016$ & $0.006 - 0.010$ & $0.006 - 0.013$\\
         Acceptance & $0.001 - 0.013$ & $0.003 - 0.015$ & $0.001 - 0.019$ & $0.004 - 0.019$ & $0.005 - 0.010$ & $0.005 - 0.025$\\
\hline
\multicolumn{7}{c}{\vthree}\\
\hline
Total syst. uncert. & $0.0015 - 0.0162$ & $0.0026 - 0.0070$ & $0.0026 - 0.0083$ & $0.0031 - 0.0064$ & $0.0029 - 0.0077$ & $0.0027 - 0.0135$\\
\hline
                PID & $0.0004 - 0.0073$ & $0.0012 - 0.0048$ & $0.0003 - 0.0046$ & $0.0010 - 0.0040$ & $0.0009 - 0.0052$ & $0.0012 - 0.0101$\\
      Track Quality & $0.0007 - 0.0205$ & $0.0018 - 0.0046$ & $0.0009 - 0.0083$ & $0.0013 - 0.0053$ & $0.0012 - 0.0066$ & $0.0012 - 0.0028$\\
          Occupancy & $0.0024 - 0.0086$ & $0.0015 - 0.0088$ & $0.0031 - 0.0063$ & $0.0027 - 0.0084$ & $0.0040 - 0.0055$ & $0.0030 - 0.0086$\\
         Acceptance & $0.0005 - 0.0205$ & $0.0027 - 0.0072$ & $0.0010 - 0.0090$ & $0.0031 - 0.0066$ & $0.0021 - 0.0088$ & $0.0024 - 0.0125$\\
\hline
\multicolumn{7}{c}{\vfour}\\
\hline
Total syst. uncert. & $0.0008 - 0.0144$ & $0.0019 - 0.0089$ & $0.0012 - 0.0073$ & $0.0018 - 0.0065$ & $0.0030 - 0.0044$ & $0.0015 - 0.0120$\\
\hline
                PID & $0.0002 - 0.0092$ & $0.0008 - 0.0059$ & $0.0004 - 0.0045$ & $0.0006 - 0.0039$ & $0.0013 - 0.0026$ & $0.0006 - 0.0090$\\
      Track Quality & $0.0004 - 0.0161$ & $0.0013 - 0.0062$ & $0.0008 - 0.0076$ & $0.0007 - 0.0024$ & $0.0022 - 0.0051$ & $0.0011 - 0.0051$\\
          Occupancy & $0.0012 - 0.0082$ & $0.0018 - 0.0104$ & $0.0015 - 0.0052$ & $0.0015 - 0.0091$ & $0.0026 - 0.0040$ & $0.0018 - 0.0078$\\
         Acceptance & $0.0004 - 0.0165$ & $0.0016 - 0.0078$ & $0.0012 - 0.0085$ & $0.0015 - 0.0055$ & $0.0031 - 0.0044$ & $0.0013 - 0.0114$\\
\hline
\end{tabular*}
\caption{Summary of the contributions to the absolute systematic 
  uncertainties for different particle species and flow coefficients 
  of different orders.  For the \pt~dependence in different regions of
  rapidity (\vone\ and \vthree: $-0.25 < \ycm < -0.15$, \vtwo\ and
  \vfour: $|\ycm| < 0.05$) and the rapidity dependence in a selected
  region of \pt\ ($1.0 < \pt < 1.5$~\gevc) the minimal and maximal
  values are given.  In addition to the total systematic
  uncertainties, also the individual contributions are shown.  These
  are due to the procedures for particle identification (''PID''), the
  quality selection criteria applied to the tracks (''Track
  Quality''), the correction for inefficiencies due to high track
  densities (''Occupancy'') and the effects of an azimuthally
  non-uniform detector acceptance (''Acceptance'').}
\label{tab:systematic_errors}
\end{table*}
%

\subsection{Correction for reconstruction inefficiencies}
\label{sect:efficiency}

%
\begin{figure*}
\begin{center}
\includegraphics[width=0.49\linewidth]{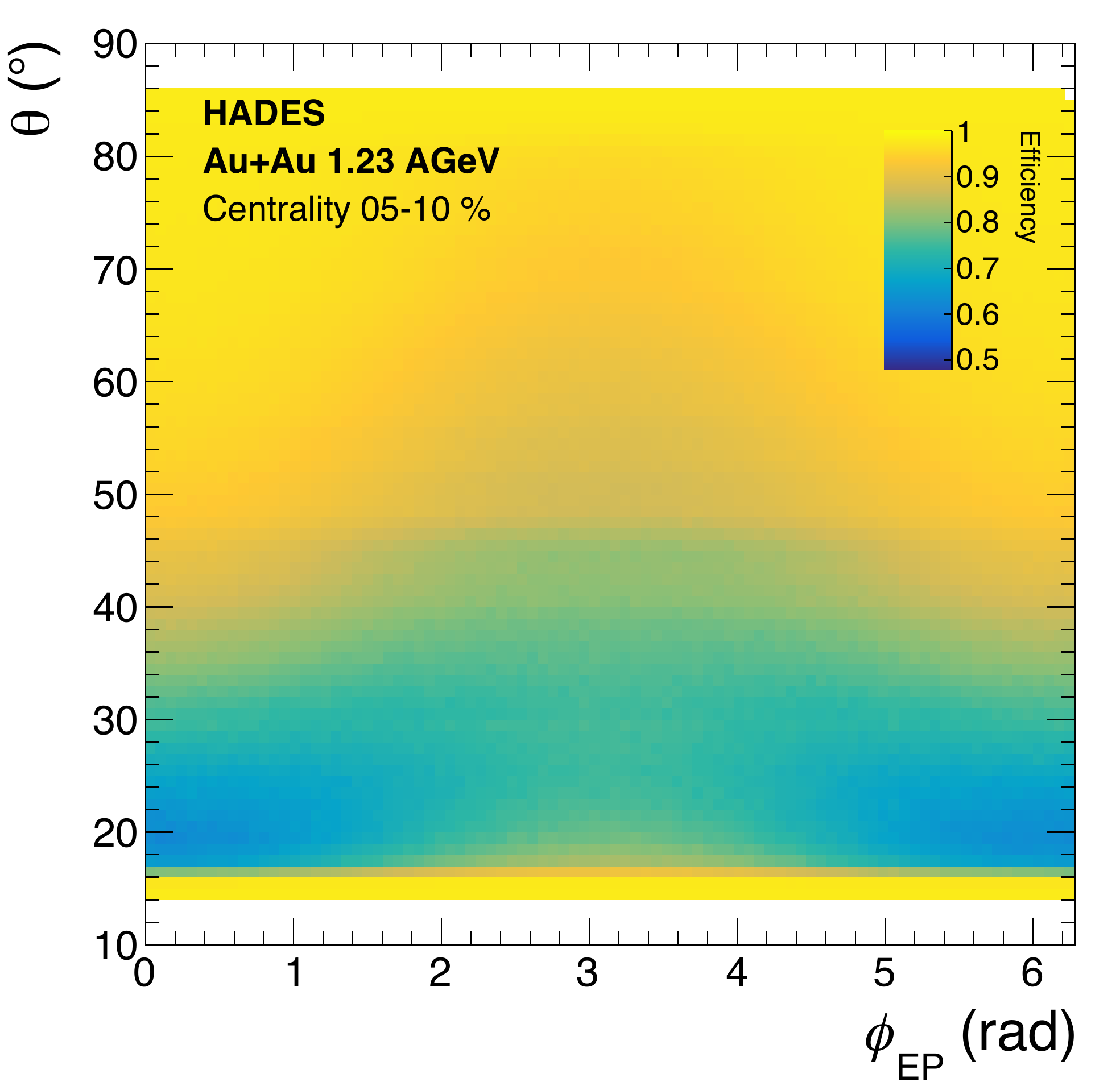}
\includegraphics[width=0.49\linewidth]{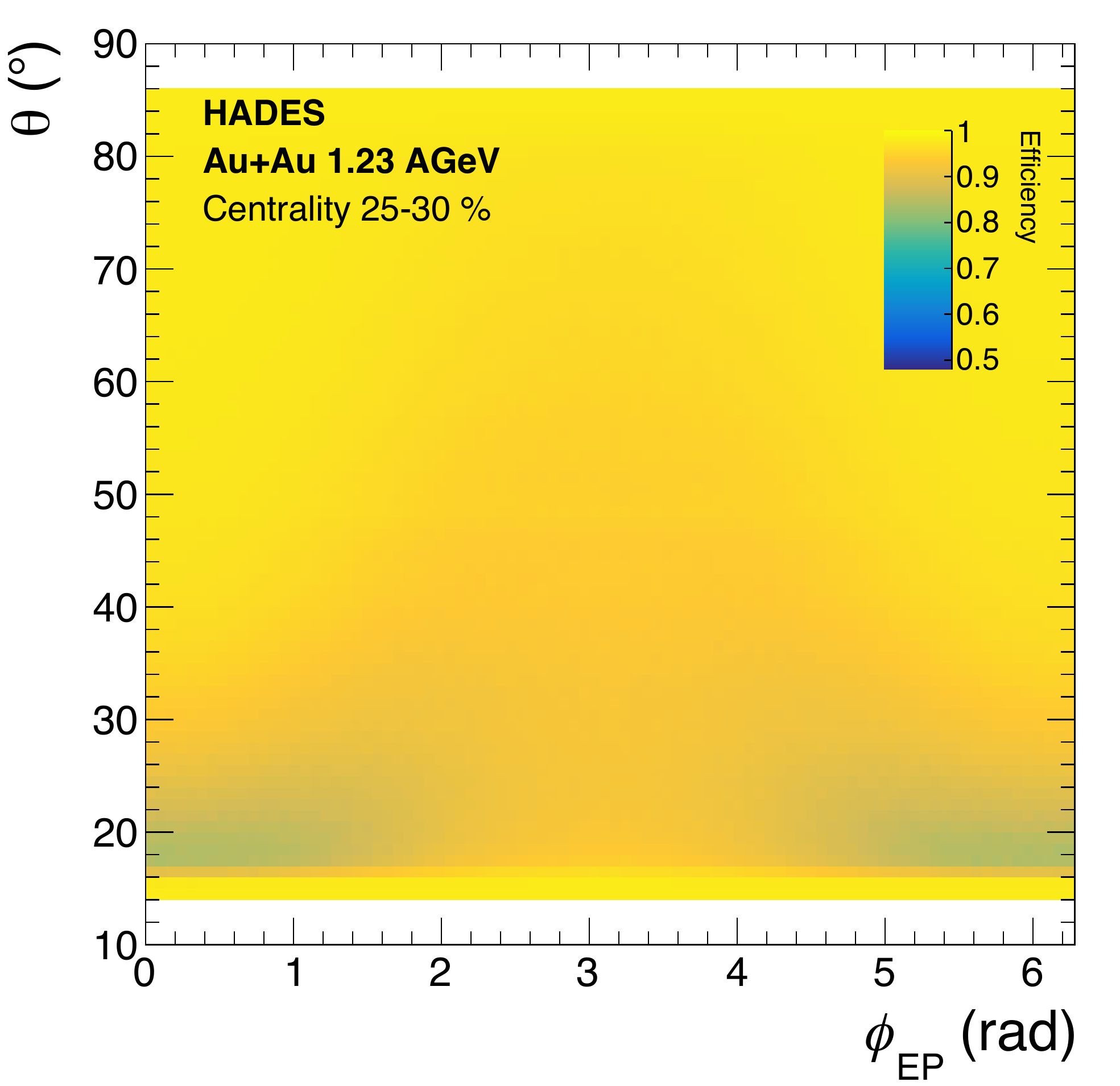}
\end{center}
\caption{The reconstruction efficiency $\epsilon(\theta,\Delta
  \phi, \textrm{cent.})$ used for all particles, shown as a
  function of the polar angle $\theta$ and of the difference between
  the azimuth angle $\phi$ and the event-plane angle $\phi_{\rb{EP}} =
  \phi - \psiepone$ for two different centrality classes (left: $5 -
  10$~\%, right: $25 - 30$~\%).}
\label{fig:efficiency}
\end{figure*}
%

In the high-multiplicity environment of Au+Au collisions the
reconstruction of tracks is affected by ambiguities in the assignment
of firing MDC drift cells to a given track.  This results in
reconstruction inefficiencies which depend on the local track
multiplicities $N_{\rb{tracks}}$.  Anisotropies in the event shape,
as caused by flow effects, will in turn generate local modulations 
of the track densities and thus of the reconstruction inefficiencies, 
which consequently distort the determination of the flow coefficients.
Therefore, any efficiency correction must also account for the track
orientation relative to the event-plane.

With the help of simulated data, generated using Geant3.21
\cite{Brun:1994aa} in combination with a detailed description of the
detector geometry and response, the multiplicity dependence of the
reconstruction efficiency $\epsilon$ was studied.  It was found that
it can be described by the following function:
\begin{equation}
\label{eq:efficiency}
  \epsilon(N_{\rb{tracks}}) = \epsilon_{\rb{max}} 
                            - c_{\epsilon} \, N_{\rb{tracks}}^{2} \, .
\end{equation}
From simulations, a maximal efficiency of $\epsilon_{\rb{max}} = 0.98$
is determined.  In the phenomenological data-driven approach used
here the parameter $c_{\epsilon}$ is adjusted such that $\vone = 0$
for $\ycm = 0$, as required by the symmetry of the reaction system.
In a next step, the average local track multiplicity $\langle
N_{\rb{tracks}}^{\rb{loc.}} \rangle$ is calculated from data in
intervals of the track polar angle $\theta$, of the difference between
its azimuth angle $\phi$ and the one of the event-plane $\phi_{\rb{EP}}
= \phi - \psiepone$ and of the event centrality.  Using these
three-dimensional matrices as input to \Eq{eq:efficiency}, relative
efficiency tables $\epsilon(\theta,\phi_{\rb{EP}}, \textrm{cent.})$ are
determined.  Examples are shown in \Fi{fig:efficiency}.  These are
then used to weight all tracks used to calculate the flow coefficients
according to
\begin{equation}
w_{\rb{eff}}(\theta,\phi_{\rb{EP}},\textrm{cent.})
   = \frac{1}{\epsilon(\theta,\phi_{\rb{EP}},\textrm{cent.})} \, .
\end{equation}

\subsection{Systematic uncertainties}
\label{sect:systematics}

%
\begin{figure*}
\begin{center}
\includegraphics[width=0.49\linewidth]{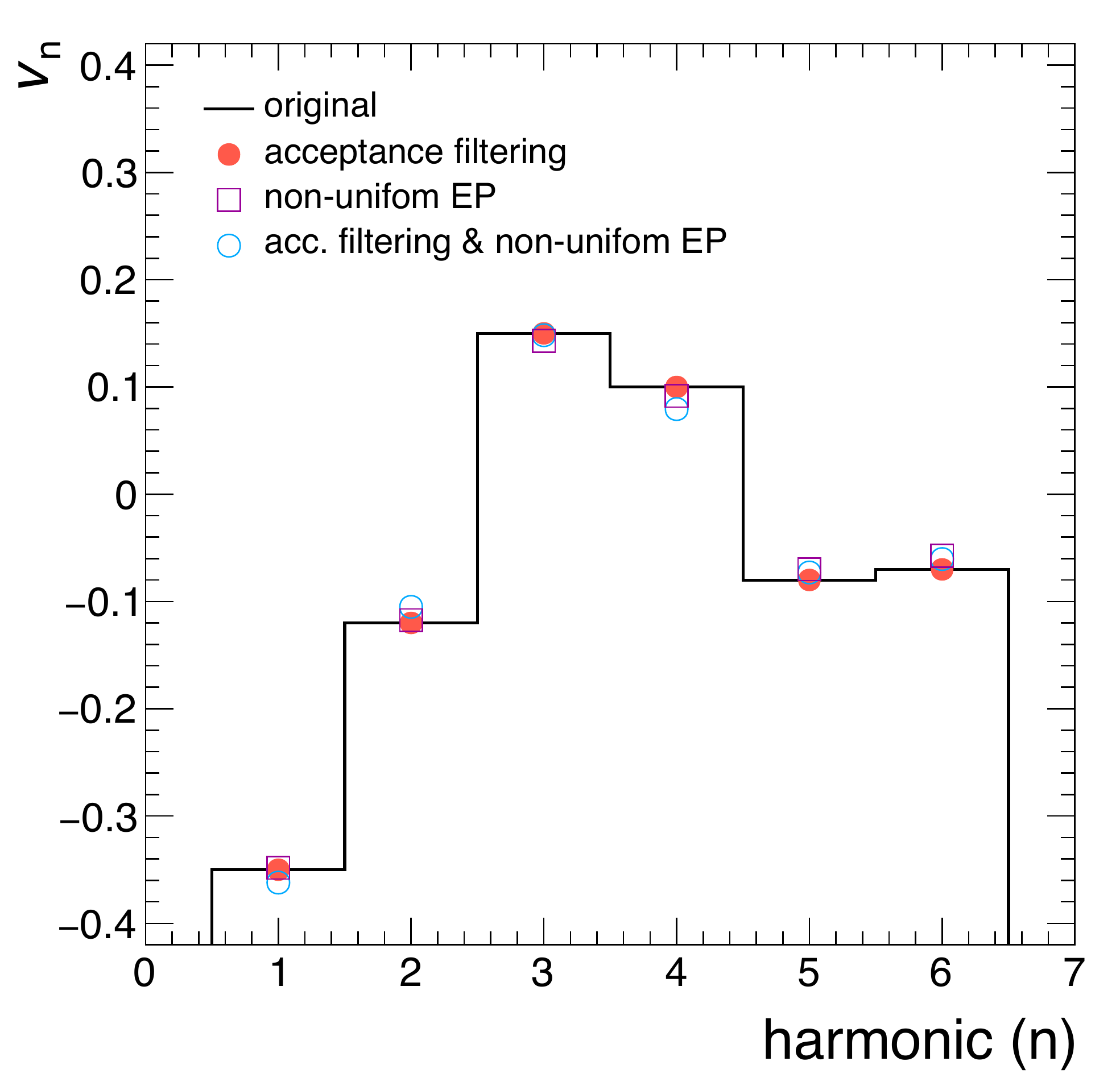}
\includegraphics[width=0.49\linewidth]{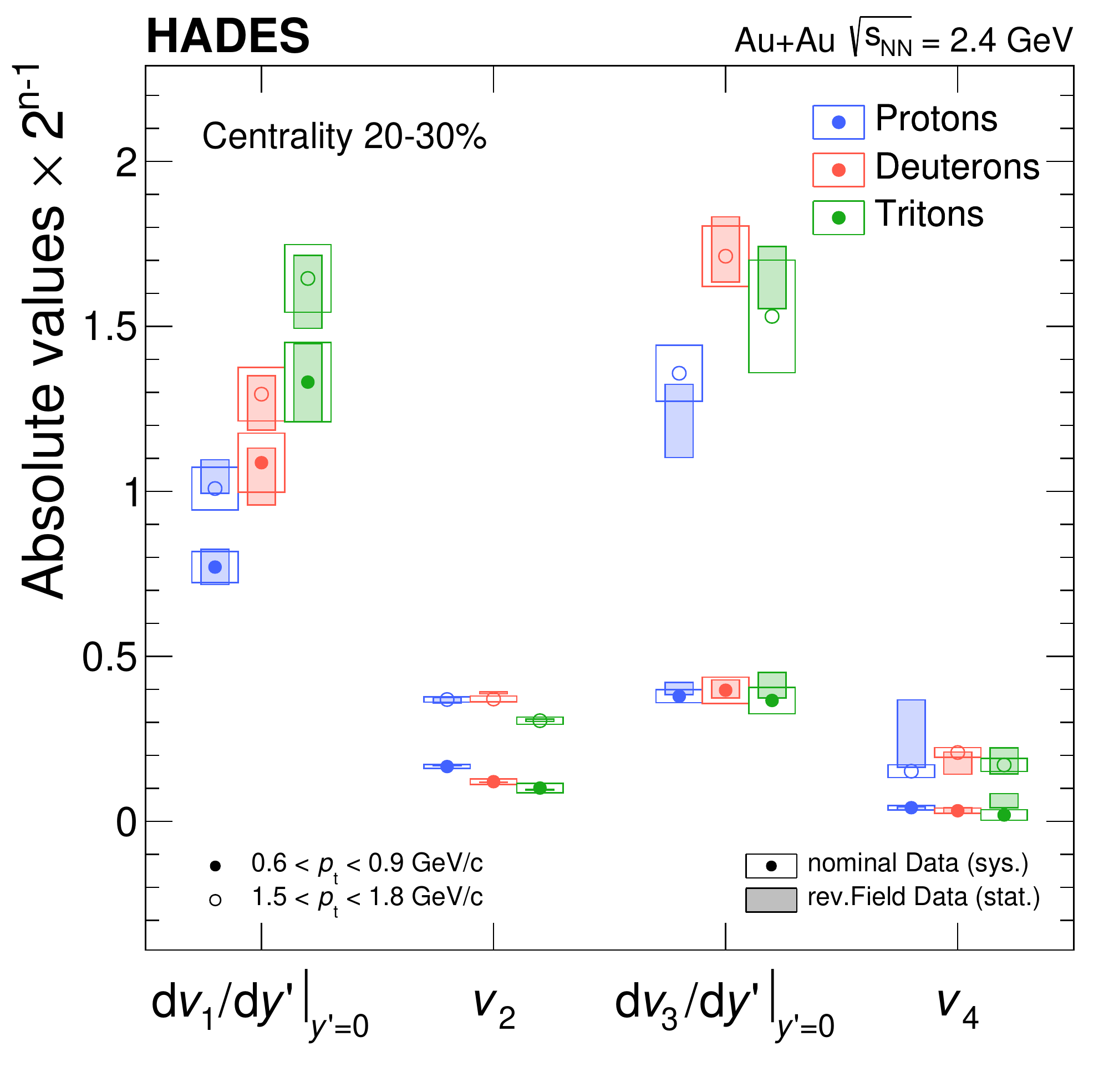}
\end{center}
\caption{Left: Result of a toy MC study to investigate the influence
  of the incomplete acceptance and a non-uniform event-plane on the
  flow coefficients of different order $n$.  The histogram represents
  the input values which are compared to the values reconstructed
  after including the different effects in the simulation.
  Right: Comparison of the flow coefficients reconstructed from
  the full data set and from the one including only data with reversed
  field polarity.  Shown are the absolute values $|\dvonemid|$,
  $|\vtwo|$, $|\dvthrmid|$ and $|\vfour|$ measured at mid-rapidity for
  two exemplary \pt~intervals and the centrality $20 - 30$~\%.  The
  data points are scaled for visibility.}
\label{fig:systematics}
\end{figure*}
%

The systematic uncertainties of the measured flow harmonics \vn\ can be
separated into global ones, i.e. those which affect all data points in
the same way, and those which depend on phase-space position. The
latter include effects of the reconstruction and selection of the
tracks, particle identification, correction procedures for
reconstruction inefficiencies and time-dependent changes of the
acceptance.  They are determined as a function of \ycm\ and \pt,
separately for each particle species, the order of the flow harmonics
\vn, and the centrality class.

The systematic uncertainties due to the track reconstruction are
estimated by varying the track selection cuts.  
Table~\ref{tab:cut-parameter} lists, in addition to the nominal
selection criteria, also two values for each cut used for the
determination of systematic effects.  Impurities in the selected
particle samples, i.e. a background of misidentified particles, will
also modify the corresponding flow result.  Their contribution to the
systematic uncertainty is evaluated by varying the PID selection
criteria (see also \Ta{tab:cut-parameter}).

The parameter $c_{\epsilon}$ used in the correction for multiplicity
dependent inefficiencies in \Eq{eq:efficiency} is modified relative to
its nominal value to evaluate the influence of the correction
procedure on the resulting $v_{n}$.  This variation covers all values
of $c_{\epsilon}$ which are still compatible within errors with $\vone
= 0$ at mid-rapidity.

In larger periods of the data-taking time, one sector of the MDC was
not fully operational. As this introduces an azimuthal asymmetry into
the acceptance and therefore increases the sensitivity to an imperfect
re-centring of the event-plane, it can be the cause of an additional
systematic uncertainty. This is estimated by comparing the results
obtained for a fully symmetric detector (i.e. six operational sectors)
with those for only five sectors. In addition, configurations were
analyzed with only four or even three active sectors, corresponding to
the upper or lower part of the detector.

The total systematic uncertainly is derived by independently analyzing
all different variations and then evaluating the overall distributions
of the resulting flow coefficients.  It is found that for the even
coefficients all the effects described above contribute roughly at the
same level to the point-by-point systematic uncertainties, whereas
azimuthal anisotropies, like efficiency losses in whole sectors,
dominate the systematic uncertainties of the odd flow coefficients.  A
summary of the different systematic uncertainties is given in
\Ta{tab:systematic_errors}.

In order to verify that the higher flow harmonics are not 
artificially generated by acceptance holes, a toy MC study was 
performed.  This simulation mimics corresponding effects by passing 
tracks through an acceptance filter.  This filter includes the gaps
between the sectors for the support structures and in addition one
entirely missing sector.  Furthermore, also the effect of a
non-uniform event-plane distribution was included.  No significant
differences between the input values \vn\ and the ones extracted after
filtering are observed, see left panel of \Fi{fig:systematics}.

Another systematic check is performed by analysing data that was
recorded with a reversed magnetic field setting.  In this
configuration, the bending directions of positively and negatively
charged particles are swapped such that they are measured by 
different areas in the outer two MDC layers, as well as TOF and RPC.  
No significant differences between the two settings are found, see
right panel of \Fi{fig:systematics}.  In addition, the analyses are
also performed for each day of data-taking separately, in order to
investigate whether any systematic trends appear in the course of the
whole data-taking time.  Also in this case, no deviations beyond the 
systematic uncertainty are observed.

Residual systematic effects can also be assessed by investigating
whether the Fourier decomposition of the azimuthal particle
distributions contains sine terms in addition to the cosine terms
in \Eq{eq:fourier}.  These are found to be of smaller or similar
magnitude than the systematic uncertainties estimated via the methods
discussed above.  Therefore, no additional systematic uncertainty is
assigned due to these differences.

The main contribution to the global systematic uncertainty arises
from the event-plane resolution.  This is mainly caused by so-called
``non-flow'' correlations which can distort the event-plane 
determination.  The magnitude of these systematic effects is 
evaluated using the three-sub-event method, i.e. by determining the 
event-plane resolution for combinations of different sub-events 
separated in rapidity.  It is found to be below 5~\% for the 
centralities $10 - 40$~\% \cite{Mamaev:2020qom}.
%

%
\begin{figure*}
\begin{center}
\includegraphics[width=0.9\textwidth]{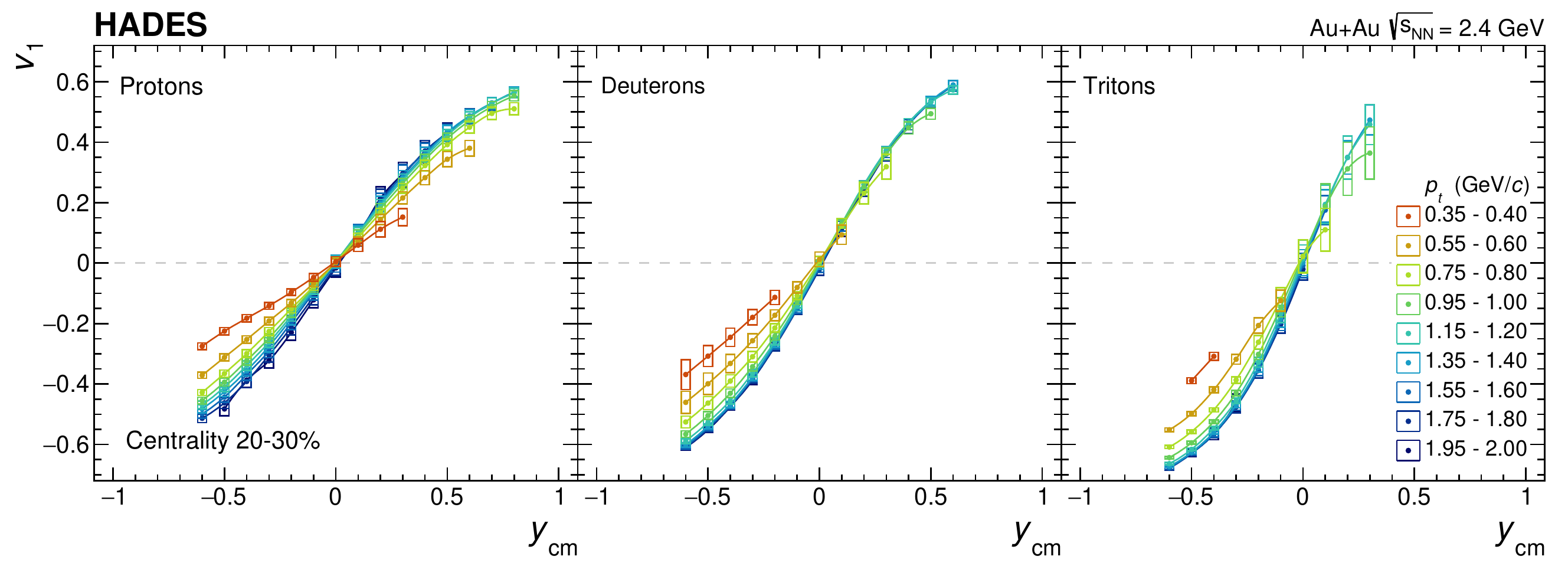}
\includegraphics[width=0.9\textwidth]{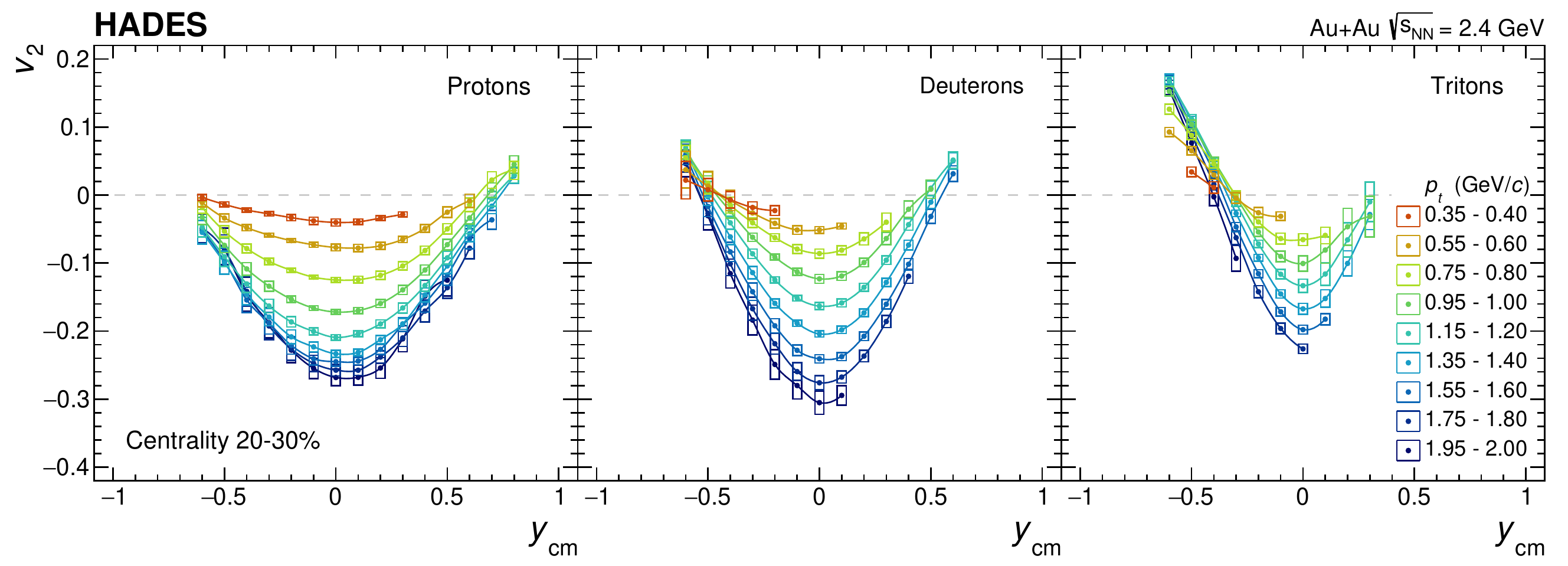}
\includegraphics[width=0.9\textwidth]{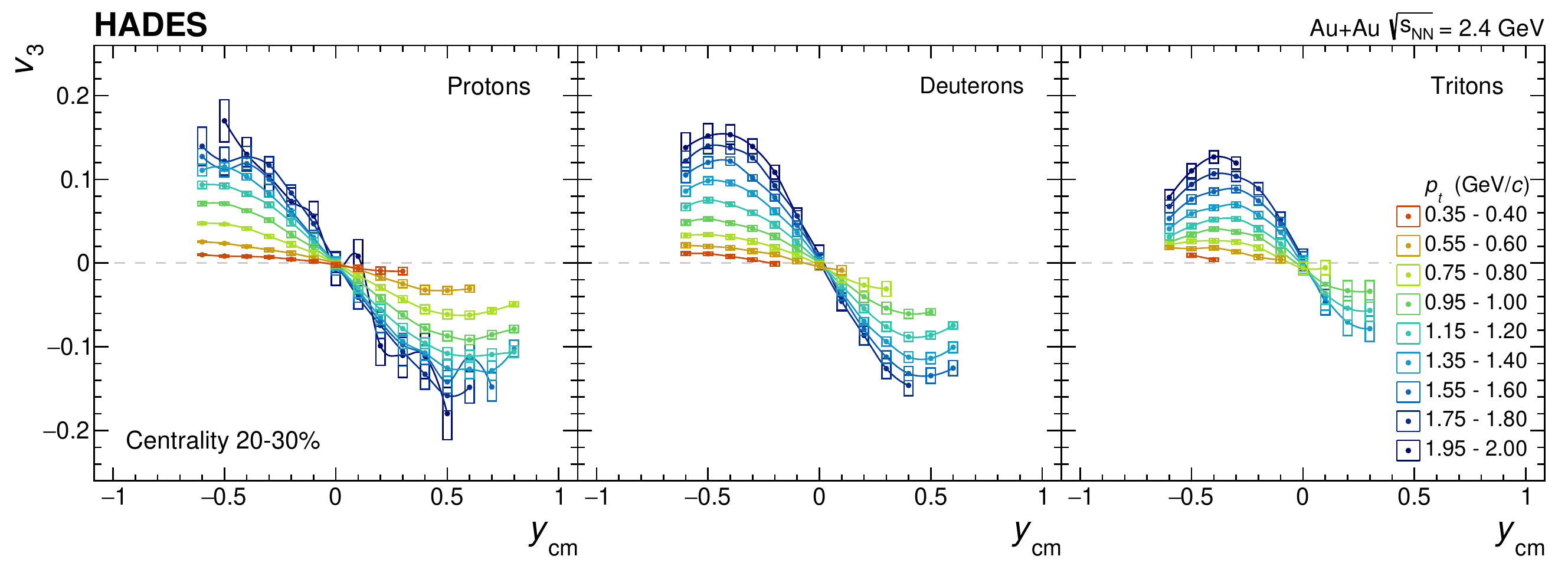}
\includegraphics[width=0.9\textwidth]{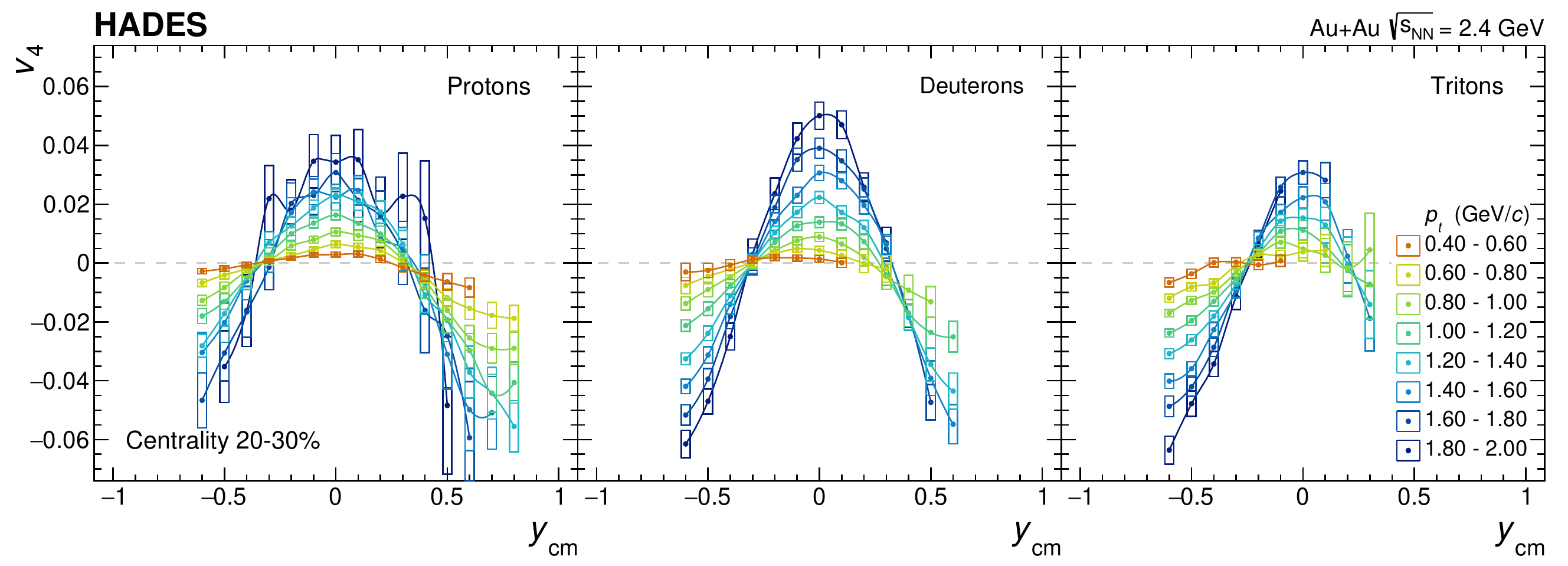}
\end{center}
\caption{The flow coefficients \vone, \vtwo, \vthree, and \vfour\
  (from top to bottom panels) of protons, deuterons and tritons (from
  left to right panels) in semi-central ($20 - 30$~\%) Au+Au
  collisions at $\sqrtsnn = 2.4$~GeV as a function of the
  centre-of-mass rapidity \ycm\ in transverse momentum intervals of
  50~\mevc\ width.  Systematic uncertainties are displayed as boxes.
  Lines are to guide the eye.}
\label{fig:vn_pdt_ycm_2030Cent}
\end{figure*}
%

%
\begin{figure*}
\begin{center}
\includegraphics[width=0.9\textwidth]{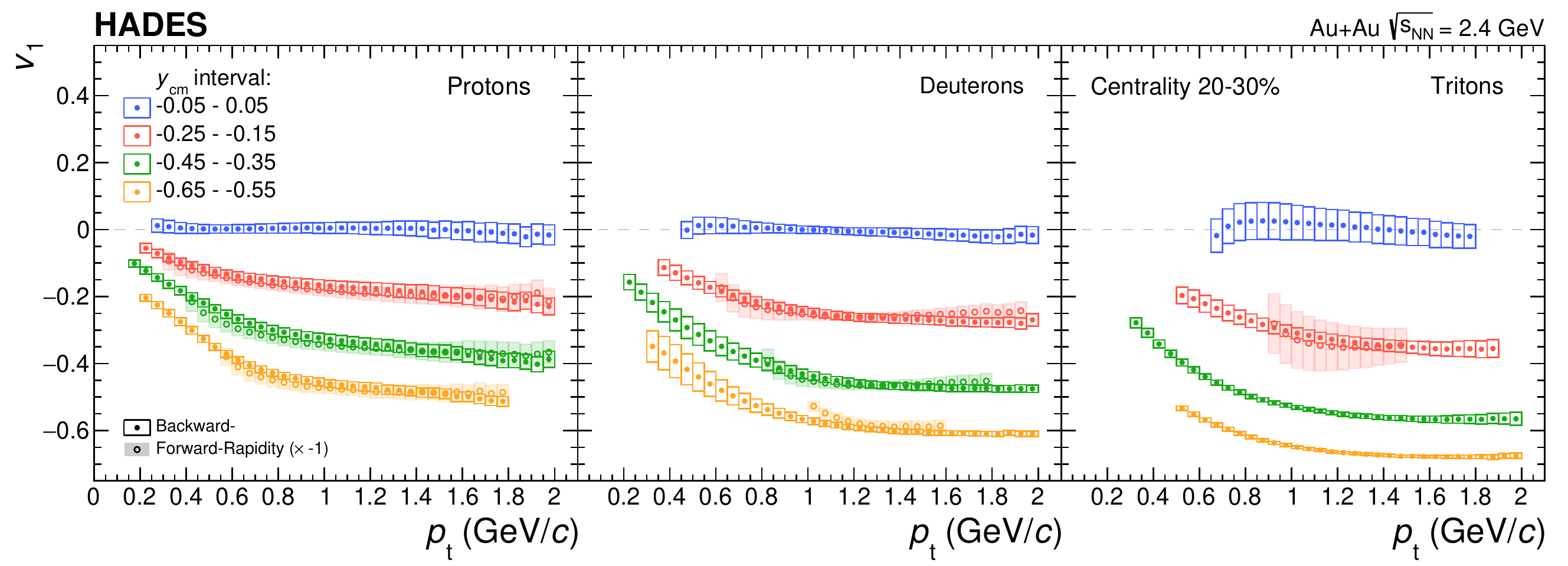}
\includegraphics[width=0.9\textwidth]{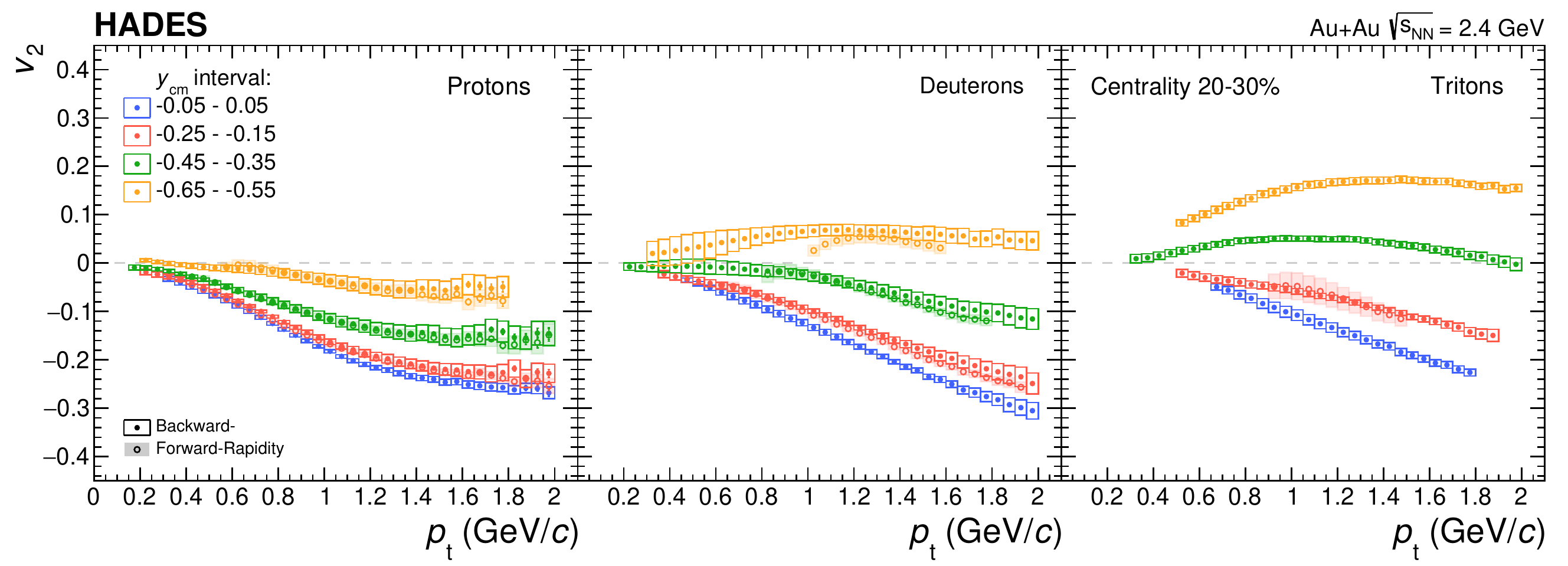}
\includegraphics[width=0.9\textwidth]{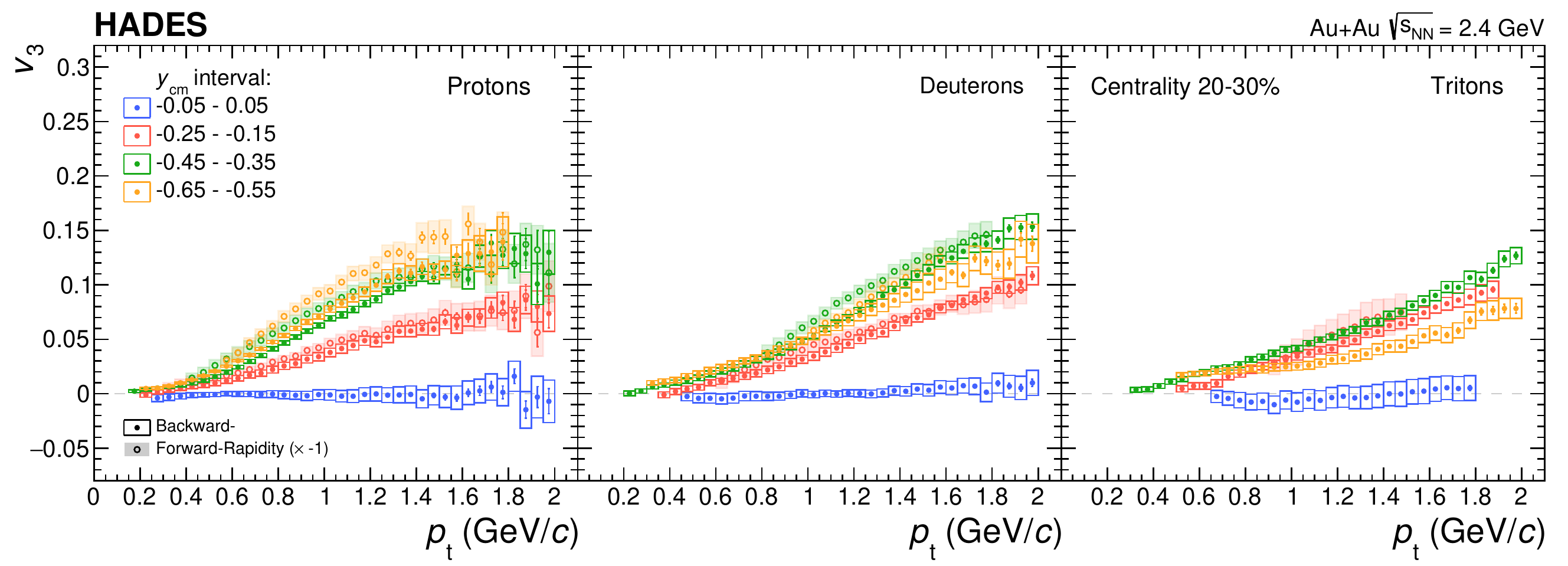}
\includegraphics[width=0.9\textwidth]{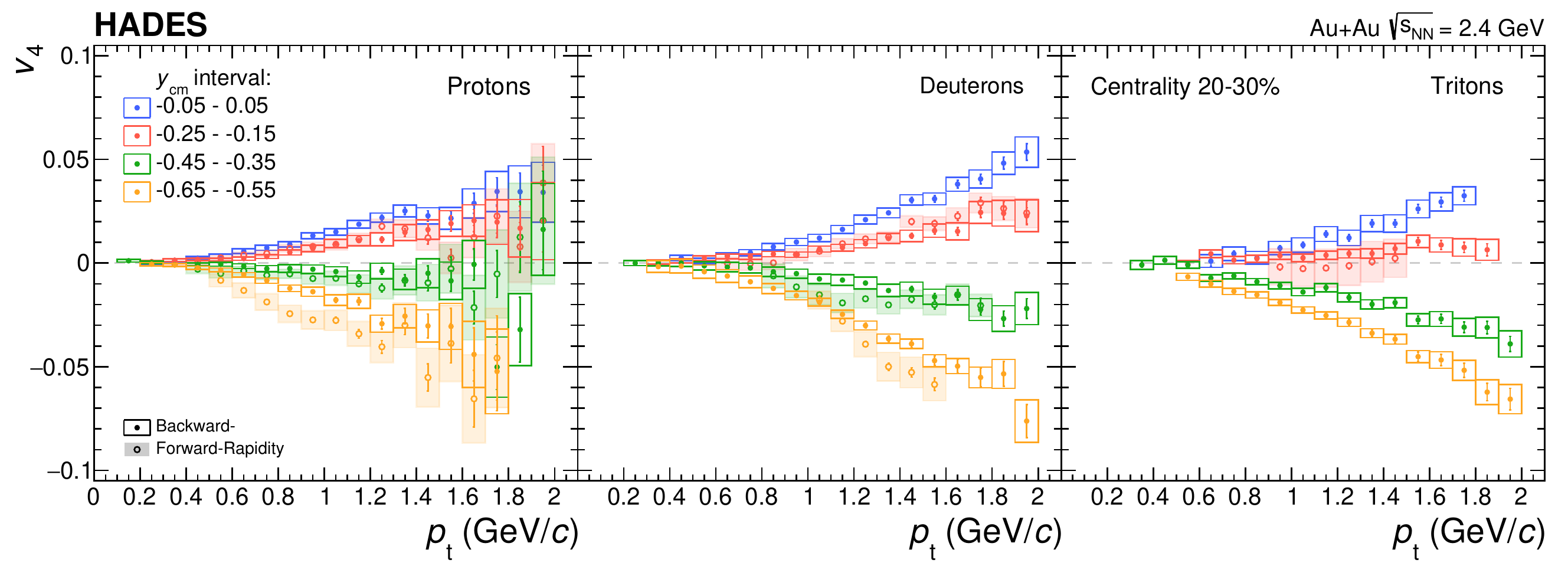}
\end{center}
\caption{The flow coefficients \vone, \vtwo, \vthree, and \vfour\
  (from top to bottom panels) of protons, deuterons and tritons (from
  left to right panels) in semi-central ($20 - 30$~\%) Au+Au
  collisions at $\sqrtsnn = 2.4$~GeV as a function of \pt\ in several
  rapidity intervals chosen symmetrically around mid-rapidity.  The
  values measured in the forward hemisphere (open symbols) have been
  multiplied by $-1$.  Systematic uncertainties are displayed as
  boxes.  Lines are to guide the eye.}
\label{fig:vn_pdt_pt_2030Cent}
\end{figure*}
%

%
\section{Flow coefficients}
\label{sect:vn}

\subsection{Directed flow: \vone}
\label{sect:vone}

Figures~\ref{fig:vn_pdt_ycm_2030Cent} and \ref{fig:vn_pdt_pt_2030Cent}
present in the uppermost row an over\-view on the directed flow
coefficient \vone\ measured for protons, deuterons and tritons in
various \pt\ and \ycm\ intervals in semi-central ($20 - 30$~\%) Au+Au
collisions.  While \vone\ of protons is consistent with zero at
mid-rapidity, it rises towards forward and decreases towards backward
rapidities (see left top panel of \Fi{fig:vn_pdt_ycm_2030Cent}).  This
rapidity dependence is stronger at higher than at lower transverse
momenta.  The \pt~dependence of the proton \vone\ is shown in the left
top panel of \Fi{fig:vn_pdt_pt_2030Cent} for four exemplary rapidity
intervals.  Its absolute value exhibits an almost linear rapid rise in
the region $\pt < 0.6$~\gevc\ and then increases only moderately or
even saturates for $\pt > 1$~\gevc.  A comparison of the absolute
\vone\ values measured in forward and backward rapidity intervals,
chosen symmetrically around mid-rapidity, results in an agreement well
within systematic errors.

A \pt\ and \ycm\ dependence similar in shape is observed for the
\vone\ of deuterons and tritons (middle and right panels in the top
row of \Fis{fig:vn_pdt_ycm_2030Cent}{fig:vn_pdt_pt_2030Cent}).
However, there are quantitative differences, namely that the
saturation behaviour sets in at higher \pt~values (above $\pt \approx
1.2$~\gevc\ for deuterons and $\pt \approx 1.4$~\gevc\ for tritons)
and reaches higher absolute values of \vone\
(e.g. $|\vone^{\rb{prot.}}| \approx 0.50$, $|\vone^{\rb{deut.}}|
\approx 0.60$ and $|\vone^{\rb{trit.}}| \approx 0.68$ for the $|\ycm|$
interval $0.55 - 0.65$).  This also implies that the dependence of
\vone\ on rapidity gets more pronounced with increasing particle mass.

\subsection{Elliptic flow: \vtwo}
\label{sect:vtwo}

Figures~\ref{fig:vn_pdt_ycm_2030Cent} and \ref{fig:vn_pdt_pt_2030Cent}
show in the second row a compilation of \vtwo\ values for protons,
deuterons and tritons as a function of \pt\ and \ycm.  Their rapidity
dependence is opposite to \vone, i.e. the absolute value of \vtwo\ is
largest at mid-rapidity and decreases towards forward and backward
rapidities for all investigated particles.  The \vtwo\ values around
mid-rapidity decrease continuously with \pt, and an indication for a
saturation behaviour is seen at relatively high \pt\ for protons only.
The drop with \pt\ is most pronounced for protons and gets weaker with
increasing particle mass.  Also the rapidity distribution of \vtwo\ in
a fixed \pt~interval strongly depends on the particle type.  While for
protons it reaches zero at rapidities of $|\ycm| \approx 0.70$, the
distributions for deuterons is significantly narrower, such that it
crosses zero already at $|\ycm| \approx 0.50$ and \vtwo\ changes sign
for larger centre-of-mass rapidities.  For tritons this change of sign
already happens around $|\ycm| \approx 0.35$.

The shape of the \pt~dependence for deuterons and tritons in the
rapidity region, where a positive \vtwo\ is observed, is clearly
different to the one in the regions with negative \vtwo.  In the
region $\ycm < -0.5$ ($\ycm < -0.35$), \vtwo\ rises with \pt\ for
deuterons (tritons) towards a maximum, whose position seems to move
towards higher \pt\ with decreasing \ycm, and then starts to drop
again.

%
\begin{figure*}
\begin{center}
\includegraphics[width=0.49\linewidth]{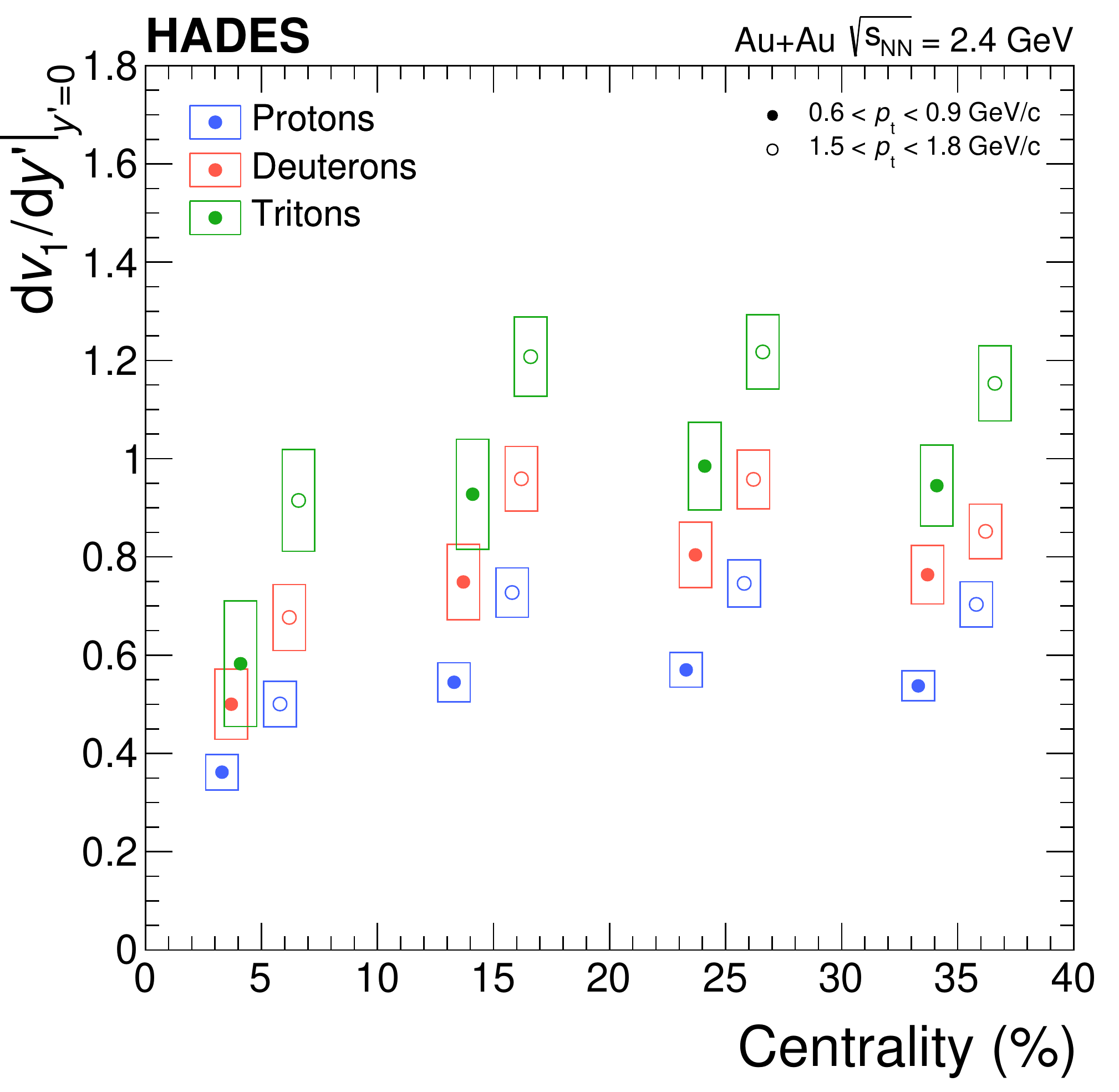}
\includegraphics[width=0.49\linewidth]{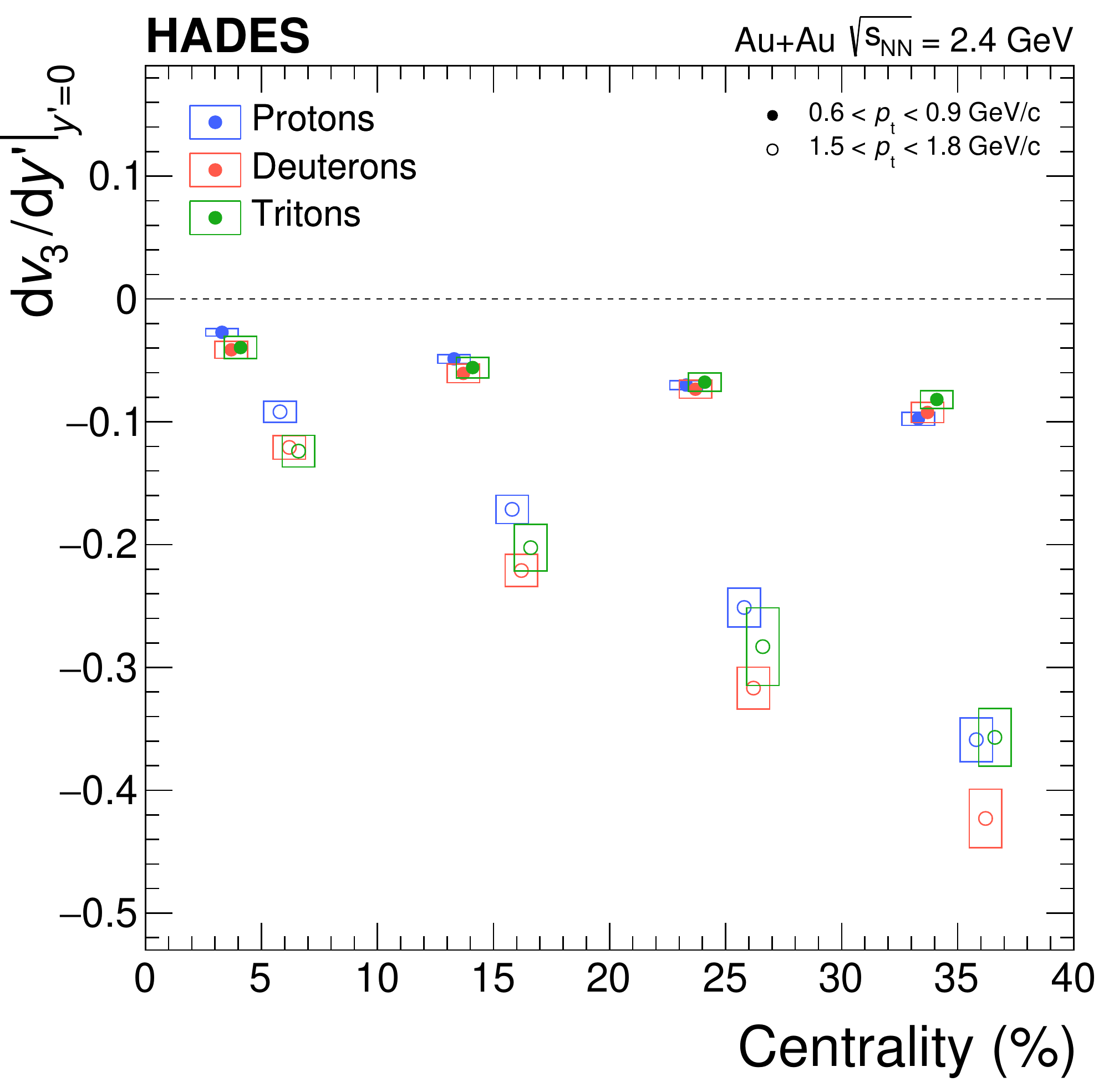}
\includegraphics[width=0.49\textwidth]{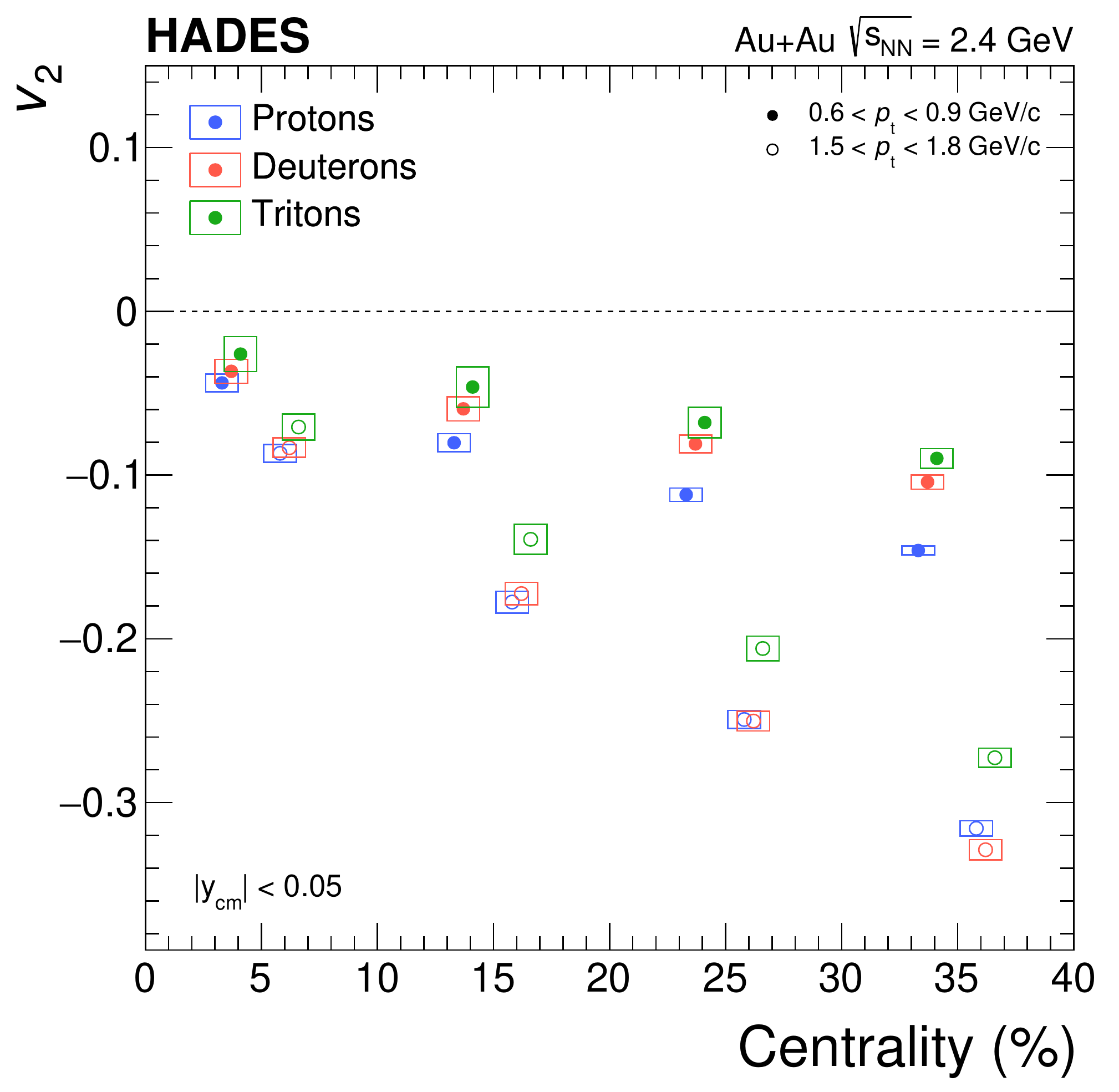}
\includegraphics[width=0.49\textwidth]{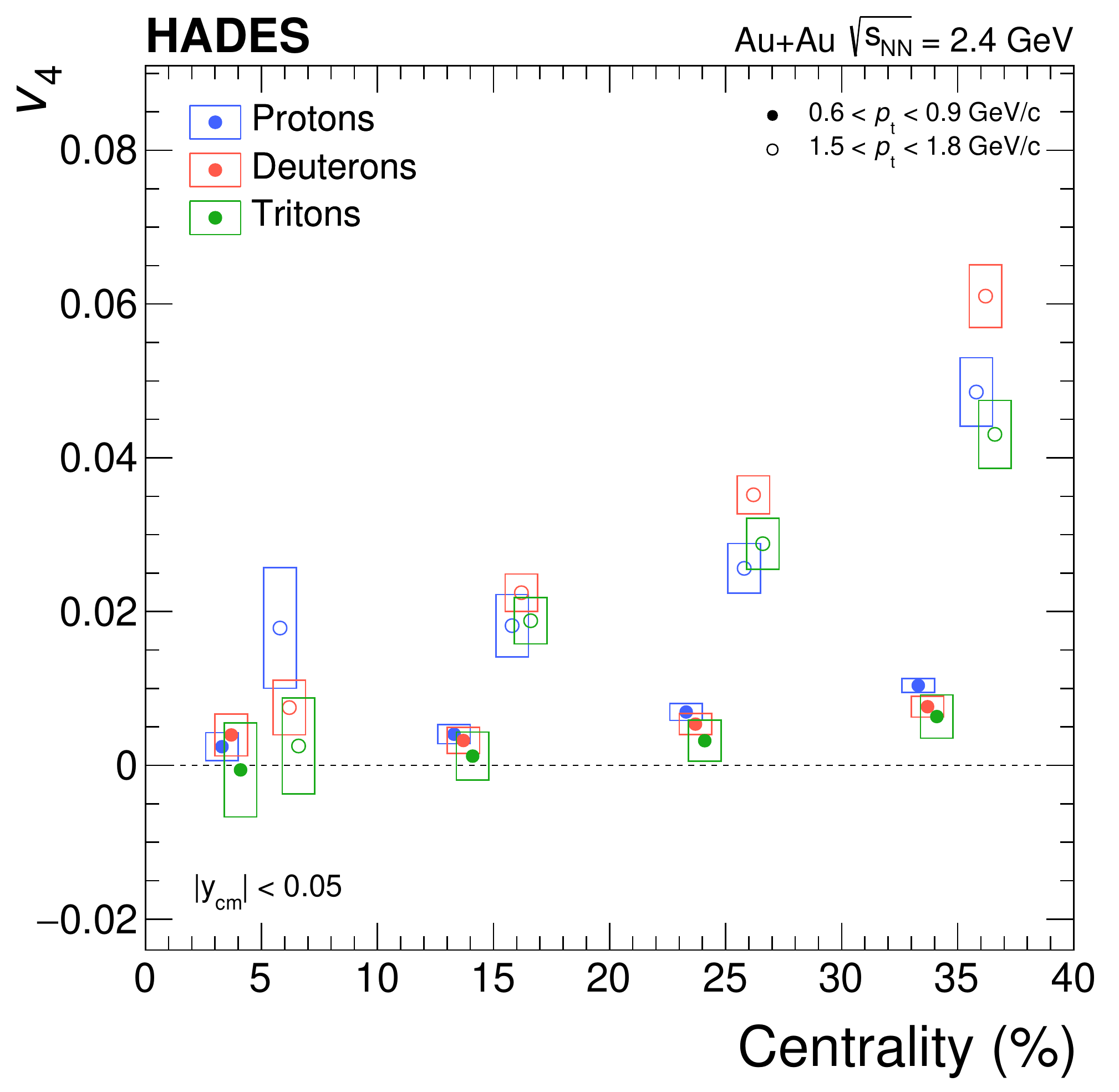}
\end{center}
\caption{Directed (\dvonemid, upper left panel), triangular (\dvthrmid,
  upper right panel), elliptic (\vtwo, lower left panel) and
  quadrangular (\vfour, lower right panel) flow of protons, deuterons
  and tritons in two transverse momentum intervals (open symbols: $0.6
  < \pt < 0.9$~\gevc\ and filled symbols: $1.5 < \pt < 1.8$~\gevc) at
  mid-rapidity in Au+Au collisions at $\sqrtsnn = 2.4$~GeV for four
  centrality classes.  Systematic uncertainties are displayed as boxes.}
\label{fig:v1-v4_all_cent}
\end{figure*}
%

\subsection{Higher flow harmonics: \vthree\ and \vfour}
\label{sect:vthree_vfour}

In addition to the directed and elliptic flow coefficients also higher
moments of the azimuthal distributions of particle emission relative
to the reaction plane have been extracted.  In
ref.~\cite{Adamczewski-Musch:2020iio} data on flow coefficients up to
the sixth order were presented for a limited region of phase-space.  A
systematic multi-differential analysis of higher orders over a larger
$\pt - \ycm$~range with satisfactory accuracy turned out to be
possible only for the coefficients \vthree\ and \vfour.

Figures~\ref{fig:vn_pdt_ycm_2030Cent} and \ref{fig:vn_pdt_pt_2030Cent}
exhibit in the third row the results on \vthree\ for protons,
deuterons and tritons.  The rapidity and \pt\ dependences are
similar for the three analysed particle species (see also
Fig.~2 in ref.~\cite{Adamczewski-Musch:2020iio}).  Generally, the
rapidity dependence is comparable in shape to the one of \vone,
however, the \vthree~values have opposite signs.  Taking a closer
look, one finds that the \ycm~distributions start with a steeper slope
at midrapidity than \vone\ and exhibit a turn around away from
mid-rapidity.  The positions of the corresponding maxima depend
slightly on the particle mass and are found at $|\ycm| \approx 0.5$
(protons), $\approx 0.4$ (deuterons) and $\approx 0.3$ (tritons).
Also, in distinction to \vone, no clear evidence for a saturation is
seen at high \pt\ for \vthree\ (see \Fi{fig:vn_pdt_pt_2030Cent}).

Figures~\ref{fig:vn_pdt_ycm_2030Cent} and \ref{fig:vn_pdt_pt_2030Cent}
present in the bottom row the results on \vfour\ for protons,
deuterons and tritons.  The rapidity distributions are similar in
shape to the ones measured for \vtwo\ for the corresponding particle,
but have opposite signs.  Also, they are narrower for \vfour\ than
for \vtwo\ and cross the $\vone = 0$ line at smaller values of
$|\ycm|$.  For the different particle species this is found to be at
$|\ycm| \approx 0.35$ (protons), $\approx 0.3$ (deuterons) and
$\approx 0.25$ (tritons).  The increase of the absolute \vfour~values
with \pt\ is also significantly less pronounced as in the case of
\vtwo.  Therefore, in contrast to the case of \vtwo, no saturation or
even a maximum is observed at higher values of \pt.

%
\begin{figure*}
\begin{center}
\includegraphics[width=0.48\linewidth]{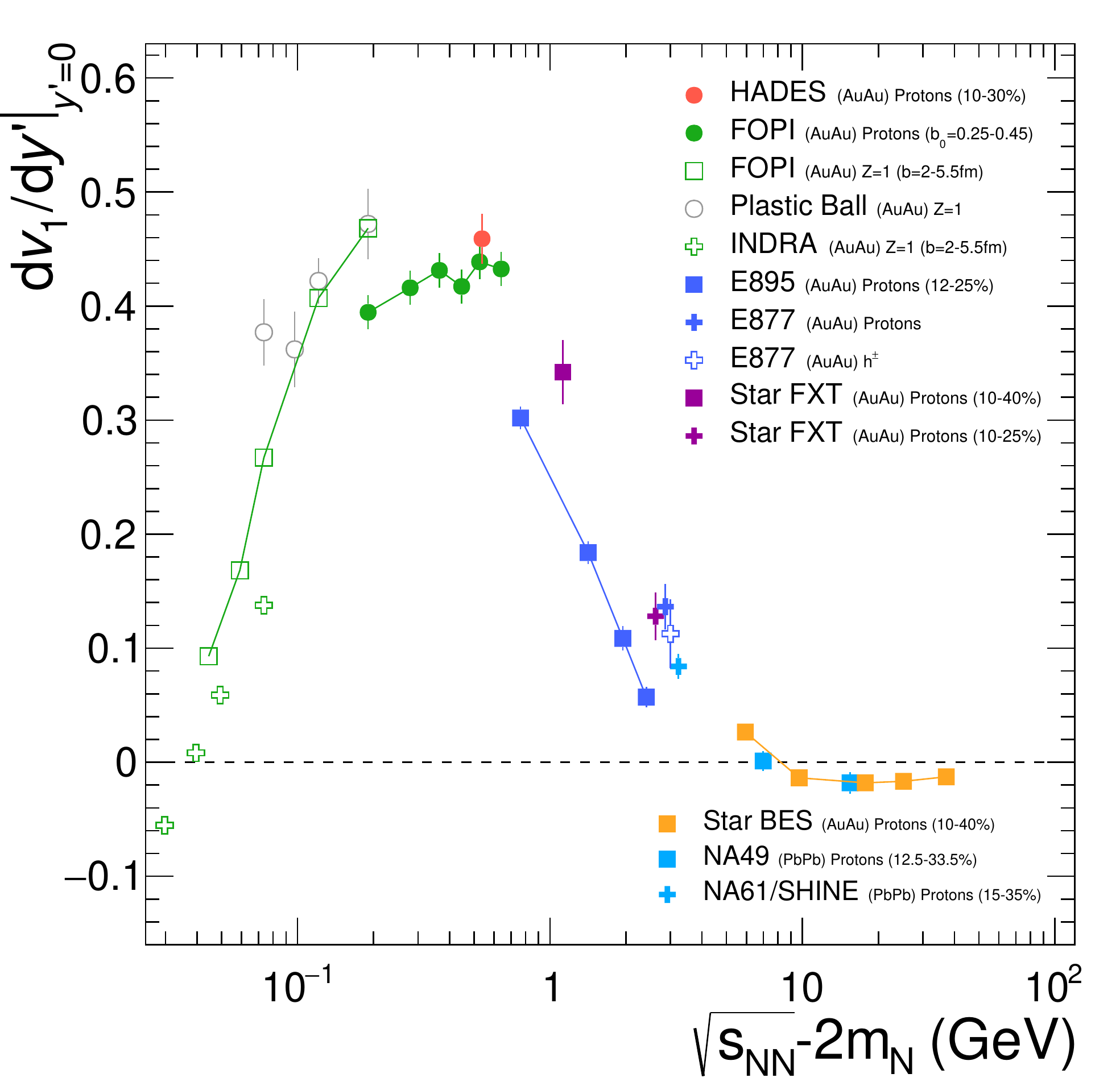}
\includegraphics[width=0.48\linewidth]{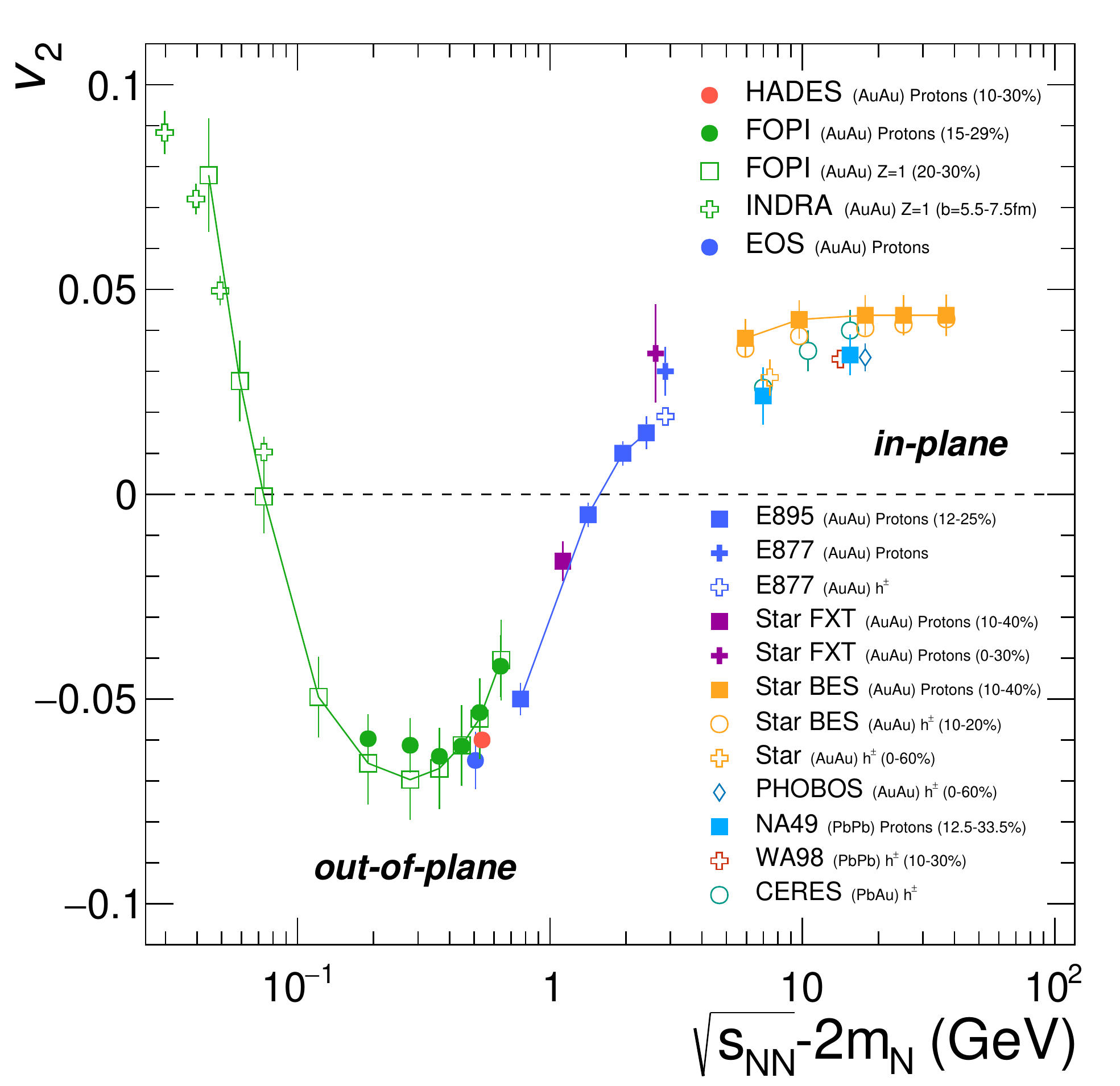}
\end{center}
\caption{Compilation of directed and elliptic flow measurements as a
  function of the subtracted centre-of-mass energy $\sqrtsnn - 2 \,
  m_{N}$.  Shown as red points are the slope of \vone\ at mid-rapidity
  (left panel), $\dvonedyp|_{y^{\prime} = 0}$, and the \pt~integrated
  \vtwo\ at mid-rapidity (right panel) for protons in Au+Au collisions
  at $\sqrtsnn = 2.4$~GeV ($10 - 30$~\% centrality).  These results
  are compared to data in the same or similar centrality ranges in
  Au+Au or Pb+Pb collisions for nuclei with $Z = 1$
  (INDRA~\cite{Andronic:2006ra},
  FOPI~\cite{Andronic:2001sw,Andronic:2004cp,Andronic:2006ra}
  Plastic Ball~\cite{Doss:1987kq,Gutbrod:1989wd}),
  for protons
  (FOPI~\cite{Andronic:2004cp,FOPI:2011aa},
  EOS/E895~\cite{Pinkenburg:1999ya,Liu:2000am},
  E877~\cite{E877:1997zjw},
  NA49~\cite{Alt:2003ab},
  STAR~\cite{Adamczyk:2014ipa,STAR:2020dav,STAR:2021ozh},
  NA61/SHINE~\cite{Kashirin:2020evw})
  and for inclusive charged particles
  (E877~\cite{Barrette:1994xr,E877:1996czs},
  CERES~\cite{Adamova:2002qx},
  WA98~\cite{Aggarwal:2004zh},
  STAR~\cite{Abelev:2009bw,Adamczyk:2012ku},
  PHOBOS~\cite{Back:2004zg}).}
\label{fig:v1v2vsEbeam}
\end{figure*}
%

%
\begin{figure}
\begin{center}
\includegraphics[width=1.0\linewidth]{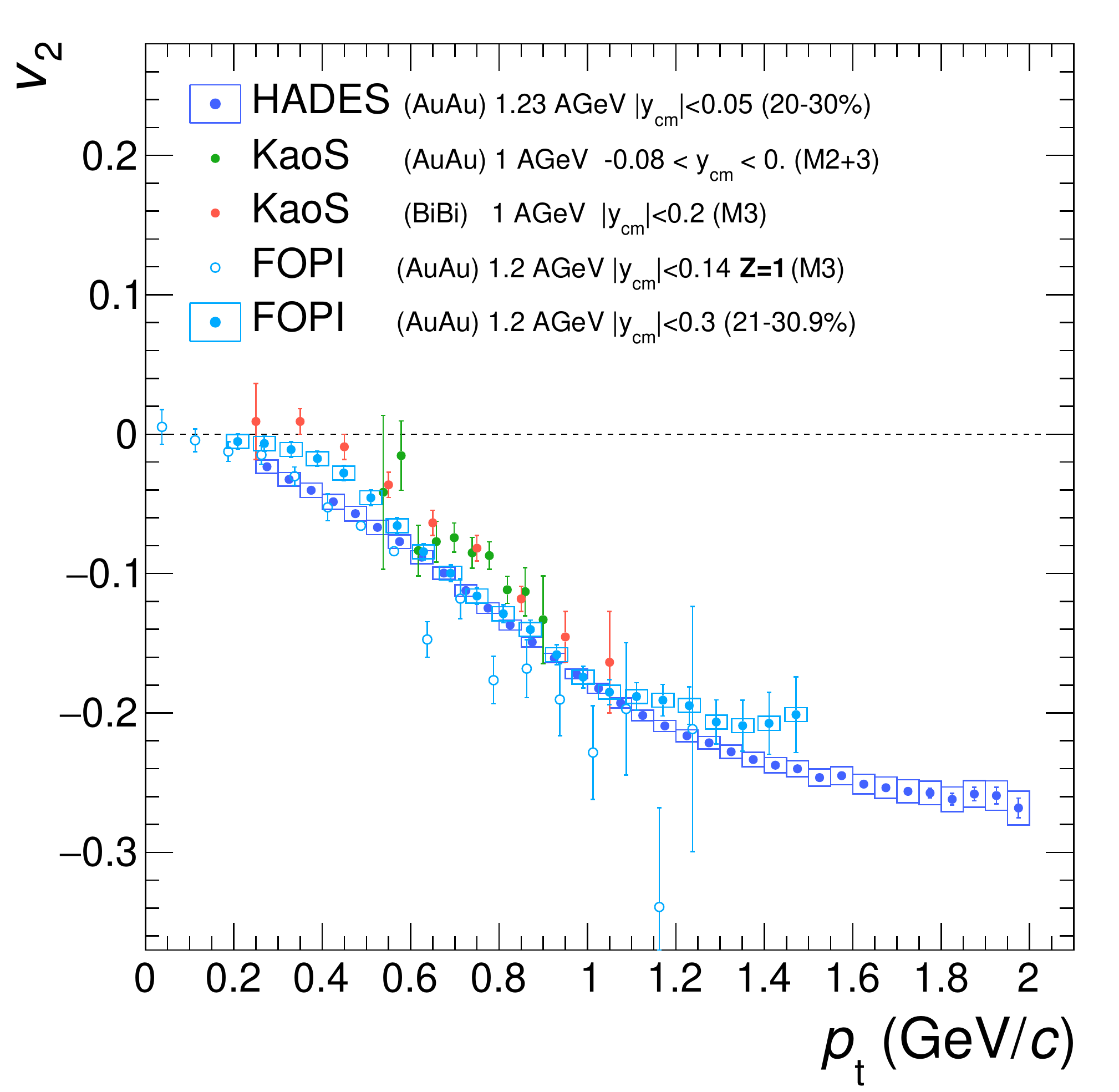}
\end{center}
\caption{Elliptic flow \vtwo\ of protons in semi-central ($20 -
  30$~\%) Au+Au collisions at $\sqrtsnn = 2.4$~GeV as a function of
  \pt\ at mid-rapidity in comparison with data measured by the
  KaoS~\cite{Brill:1996} and
  FOPI~\cite{Andronic:2004cp,FOPI:2011aa} collaborations in the same energy 
  region and for similar centrality selections.  Systematic 
  uncertainties are displayed as boxes where available.}
\label{fig:v2_compEXP_20-30}
\end{figure}
%

\subsection{Centrality dependences}
\label{sect:centrality}

The directed flow at mid-rapidity can be quantified by its slope
\dvonemid\ which is defined relative to the scaled rapidity
$y^{\prime} = \ycm/y_{\rb{mid}}$, with $y_{\rb{mid}} = 0.74$ as
mid-rapidity in the laboratory system.  It is determined as the linear
term, $\dvonemid = a_{1}$, of a cubic ansatz $\vone(y^{\prime}) =
a_{1}\,y^{\prime} + a_{3}\,y^{\prime\,3}$ which has been fitted to the
measured data points.  Similarly, the slope of \vthree\ \dvthrmid\ is
extracted. The upper panels of \Fi{fig:v1-v4_all_cent} displays the
corresponding values as determined for two different \pt~intervals
($0.6 < \pt < 0.9$~\gevc\ and $1.5 < \pt < 1.8$~\gevc) and for the
four centrality classes investigated in this analysis.  The slope of
\vone\ exhibits no significant centrality dependence for all particles
and \pt~intervals, except for the very central class where \dvonedyp\
is smaller than for the other centralities.  This is distinctly
different to the centrality dependence of the slope of \vthree, where
the absolute value $|\dvthrdyp|$ is continuously increasing with
centrality.  Also, the values are almost identical for the different
particles at all centralities, while for \dvonedyp\ a significant mass
hierarchy is observed.

Similar to the triangular flow, also \vtwo\ and \vfour\ at mid-rapidity
depend on the reaction centrality.  While the absolute value $|\vtwo|$
increases roughly linearly with centrality (see lower left panel of
\Fi{fig:v1-v4_all_cent}), \vfour\ exhibits a stronger centrality
dependence (see lower right panel of \Fi{fig:v1-v4_all_cent}).  The mass
hierarchy is visible for \vtwo\ and \vfour\ in the lower \pt~interval
for all centrality classes.  In the higher \pt~region, however, only
\vtwo\ of tritons is different from the one of protons and deuterons,
while the \vfour\ values do not exhibit any systematic ordering.
%

%
\section{Comparison to world data}
\label{sect:comparison}

The energy dependence of directed and elliptic flow at mid-rapidity
and integrated over transverse momentum is presented in
\Fi{fig:v1v2vsEbeam}.  The slope of \vone\ is shown in the left panel
and \vtwo\ at mid-rapidity in the right panel for published data
together with our data point.  Due to the lack of other measurements
in the low-energy region, a similar comparison of \vthree\ and \vfour\
can not be done.

The slope of \vone\ is characterized by a strong beam energy
dependence in the region $\ebeam/A \lesssim 10$~GeV.  While
\dvonemid\ is negative below $\ebeam/A \approx$ \linebreak $0.1$~GeV,
it is positive at higher energies and rises rapidly towards a maximum
at around $\ebeam/A \approx 1$~GeV and then drops to values close to
zero at higher beam energies.  The Au+Au reactions at $\sqrtsnn =
2.4$~GeV investigated here are in the region of maximum observable
directed flow.  A good agreement between the result of this analysis
and data measured by the FOPI collaboration \cite{FOPI:2011aa} is
found.  The characteristic energy dependence is the result of the
interplay between two effects: (i) an increasing pressure of the
fireball created in the overlap region of the reaction system, which
can push the light spectator nucleons into the reaction plane; (ii) a
decreasing passage time of the colliding nuclei, which reduces the
pressure transfer onto the light nuclei at higher energies.

Also \vtwo\ at mid-rapidity exhibits a very distinct energy dependence.  
In the region $0.1 \lesssim \ebeam/A \lesssim 5$~GeV \vtwo\ is 
negative, i.e. the particle emission is out-of-plane, as the passage 
time of the spectator matter is long enough to cause the squeeze-out
effect \cite{Reisdorf:1997fx,Stoecker:1986ci}, i.e. the fireball
pressure pushes particles preferentially into the direction which is
not shadowed by spectators.  At higher energies the passage times are
too short and particle emission is in-plane as the pressure gradients
are steepest in this direction.  The integrated \vtwo\ obtained for
Au+Au collisions at $\sqrtsnn = 2.4$~GeV in this analysis is in the
region where out-of-plane emission is still very strong. It is also
well in accordance with other measurements by
EOS~\cite{Pinkenburg:1999ya} and FOPI~\cite{Andronic:2004cp} (see
right panel of \Fi{fig:v1v2vsEbeam}).

The \pt~dependence of \vtwo\ at mid-rapidity measured by HADES is
compared with results of other experiments in the same energy region
(KaoS~\cite{Brill:1996} and FOPI~\cite{Andronic:2004cp}) in
\Fi{fig:v2_compEXP_20-30}.  Within uncertainties and considering the
slight differences of beam energies, good agreement with the other
data sets is found.  The new HADES data extend the phase-space
coverage significantly in comparison to previous measurements with
clearly improved accuracy.
%

%
\begin{figure}
\begin{center}
\includegraphics[width=1.0\linewidth]{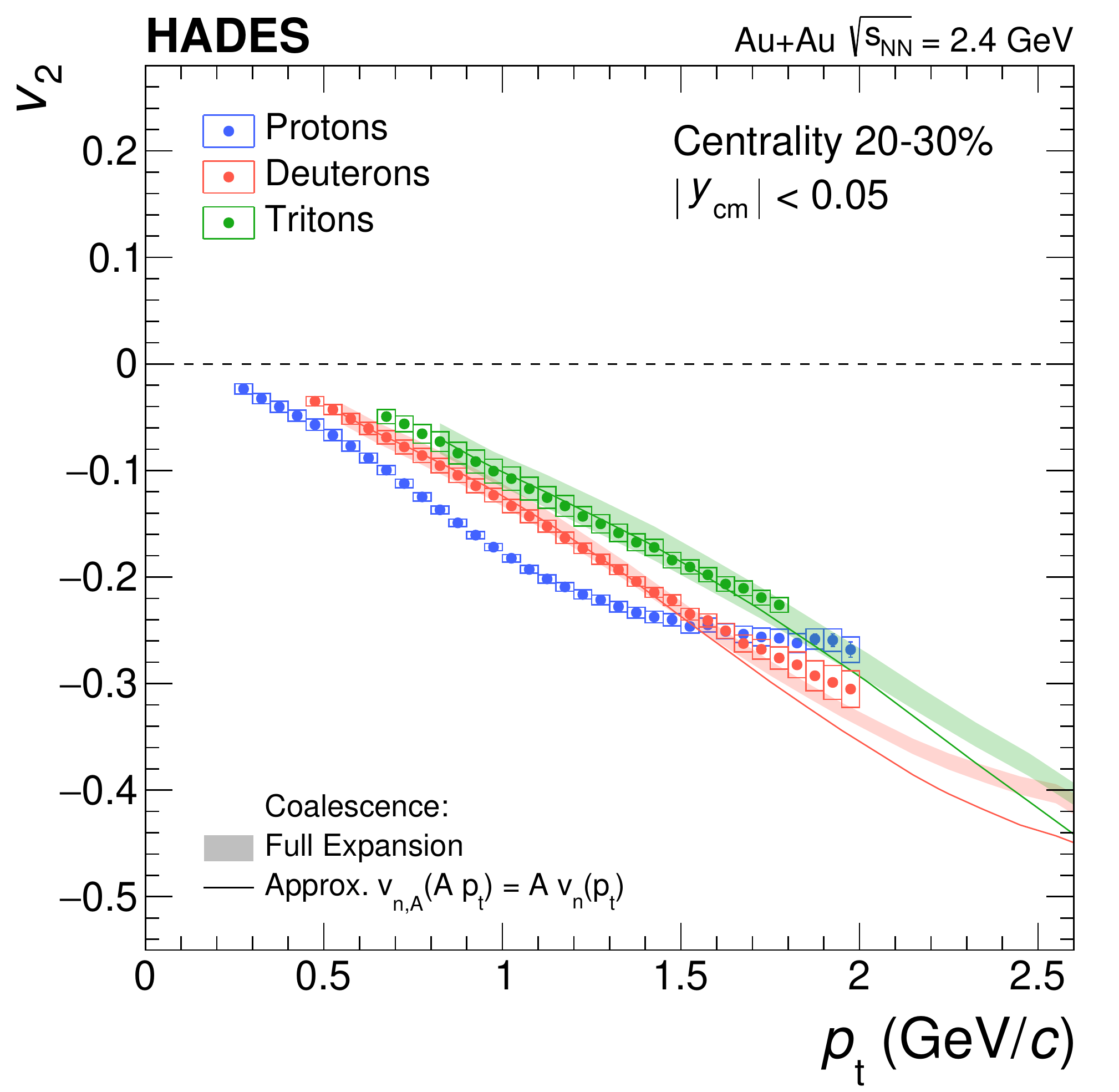}
\end{center}
\caption{Elliptic flow \vtwo\ of protons, deuterons and tritons in
  semi-central ($20 - 30$~\%) Au+Au collisions at $\sqrtsnn = 2.4$~GeV
  as a function of \pt\ at mid-rapidity ($|\ycm| < 0.05$). The solid 
  curves represent the proton distribution after scaling according to
  $v_{n,A}(A \, \pt) = A \, \vn(\pt)$. The coloured bands depict the
  results as calculated for the higher-order nucleon-coalescence
  scenario given in \Eq{eq:vn_coal_full}.  Systematic uncertainties are
  displayed as boxes.}
\label{fig:v2_all_scaling_20-30}
\end{figure}
%

%
\begin{figure}
\begin{center}
\includegraphics[width=1.0\linewidth]{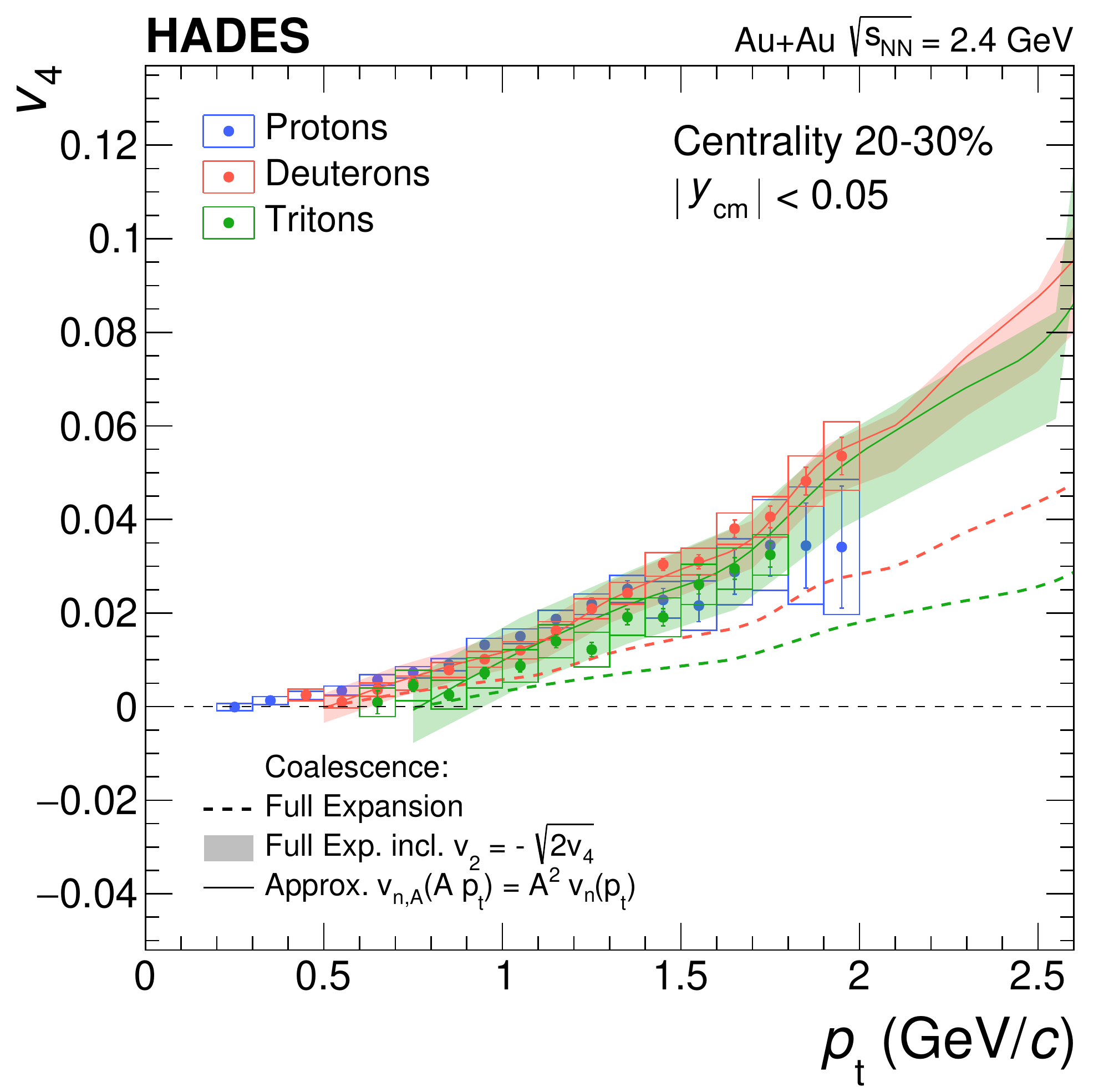}
\end{center}
\caption{Quadrangular flow \vfour\ of protons, deuterons and tritons in
    semi-central ($20 - 30$~\%) Au+Au collisions at $\sqrtsnn =
    2.4$~GeV as a function of \pt\ at mid-rapidity ($|\ycm| < 0.05$).
    The dashed curves represent the proton distribution after scaling
    according the higher order nucleon-coalescence scenario given in
    \Eq{eq:vn_coal_full}.  The coloured bands depict the results as
    calculated with \Eq{eq:v4_full_withv4v22} which includes the
    additional contribution of \vtwo\ assuming the relation $\vtwo = -
    \sqrt{2 \, \vfour}$.  The solid curves show the result for the
    approximation $v_{n,A}(A \, \pt) = A^{2} \, \vn(\pt)$.  Systematic
    uncertainties are displayed as boxes.}
\label{fig:v4_scaling_all_20-30}
\end{figure}
%

%
\begin{figure}
\begin{center}
\includegraphics[width=1.0\linewidth]{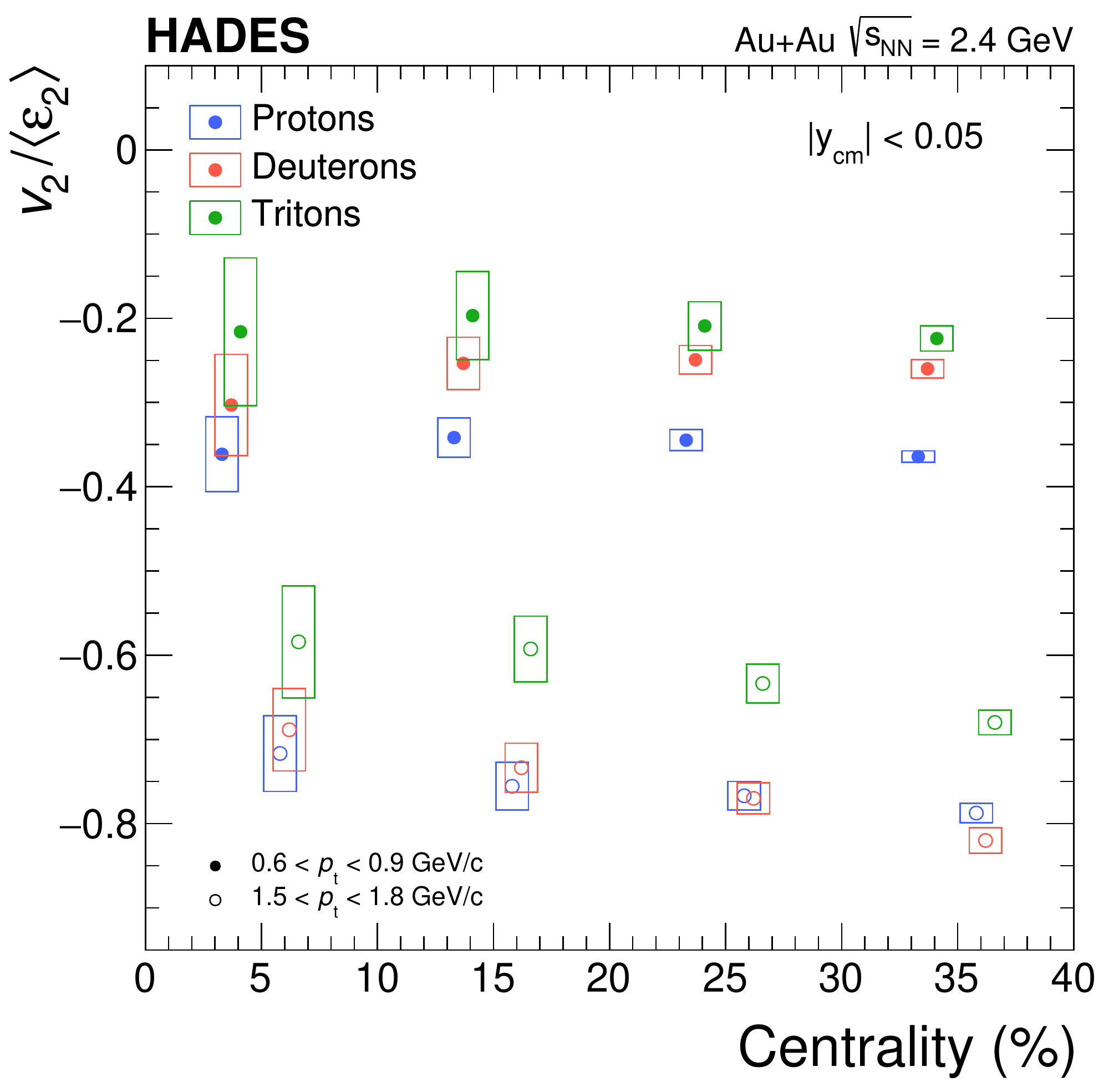}
\end{center}
\caption{Scaled elliptic flow of protons, deuterons
  and tritons in two transverse momentum intervals at mid-rapidity in
  Au+Au collisions at $\sqrtsnn = 2.4$~GeV for four centrality
  classes.  The values are divided by the second order eccentricity,
  $\vtwo / \etwoav$, as calculated within the Glauber-MC approach for
  the corresponding centrality range (see \Ta{tab:ecc2GlauberMC}).
  Systematic uncertainties are displayed as boxes.}
\label{fig:v2_eccScale}
\end{figure}
%

%
\begin{figure*}
\begin{center}
\includegraphics[width=0.48\linewidth]{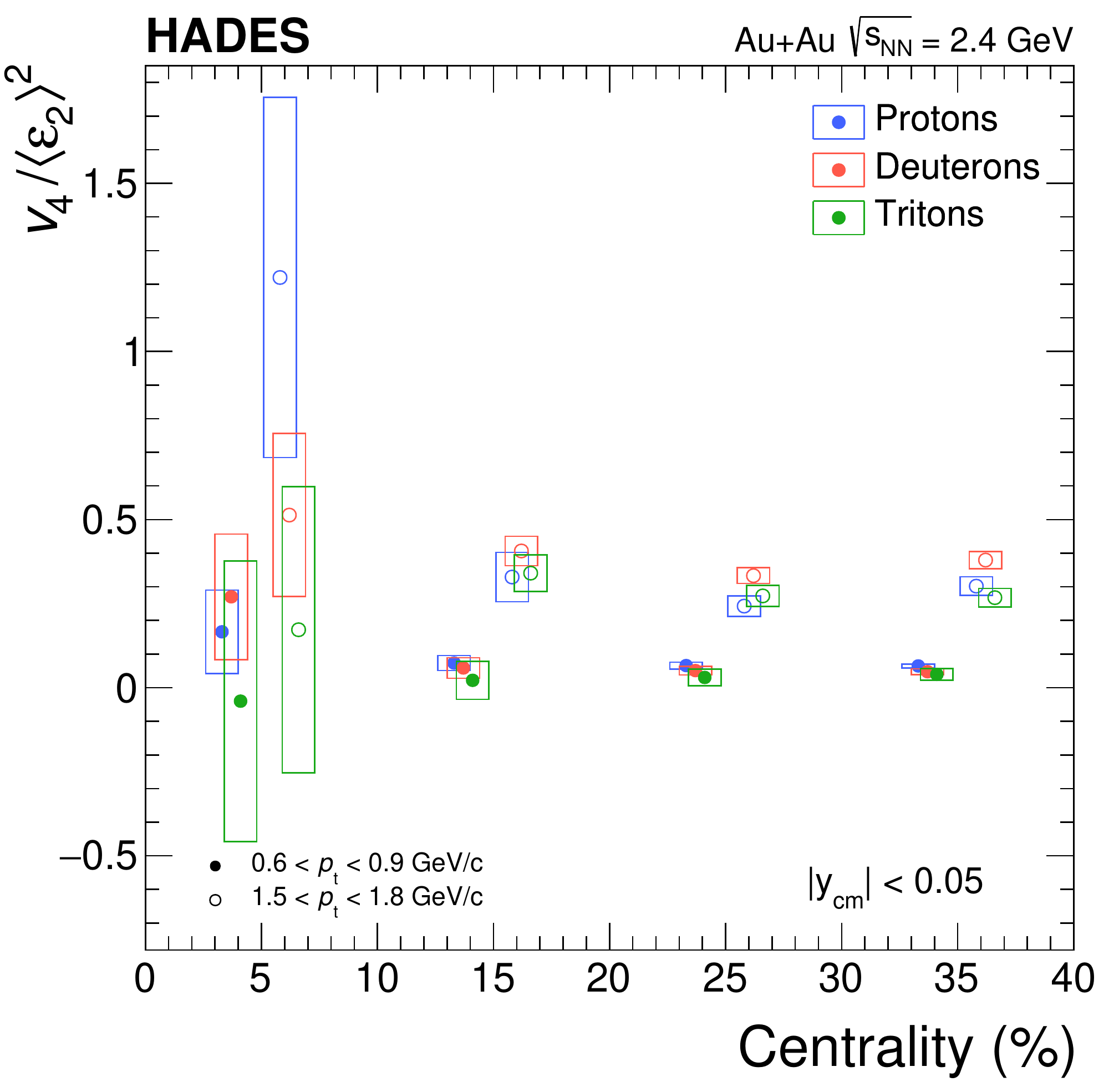}
\includegraphics[width=0.48\linewidth]{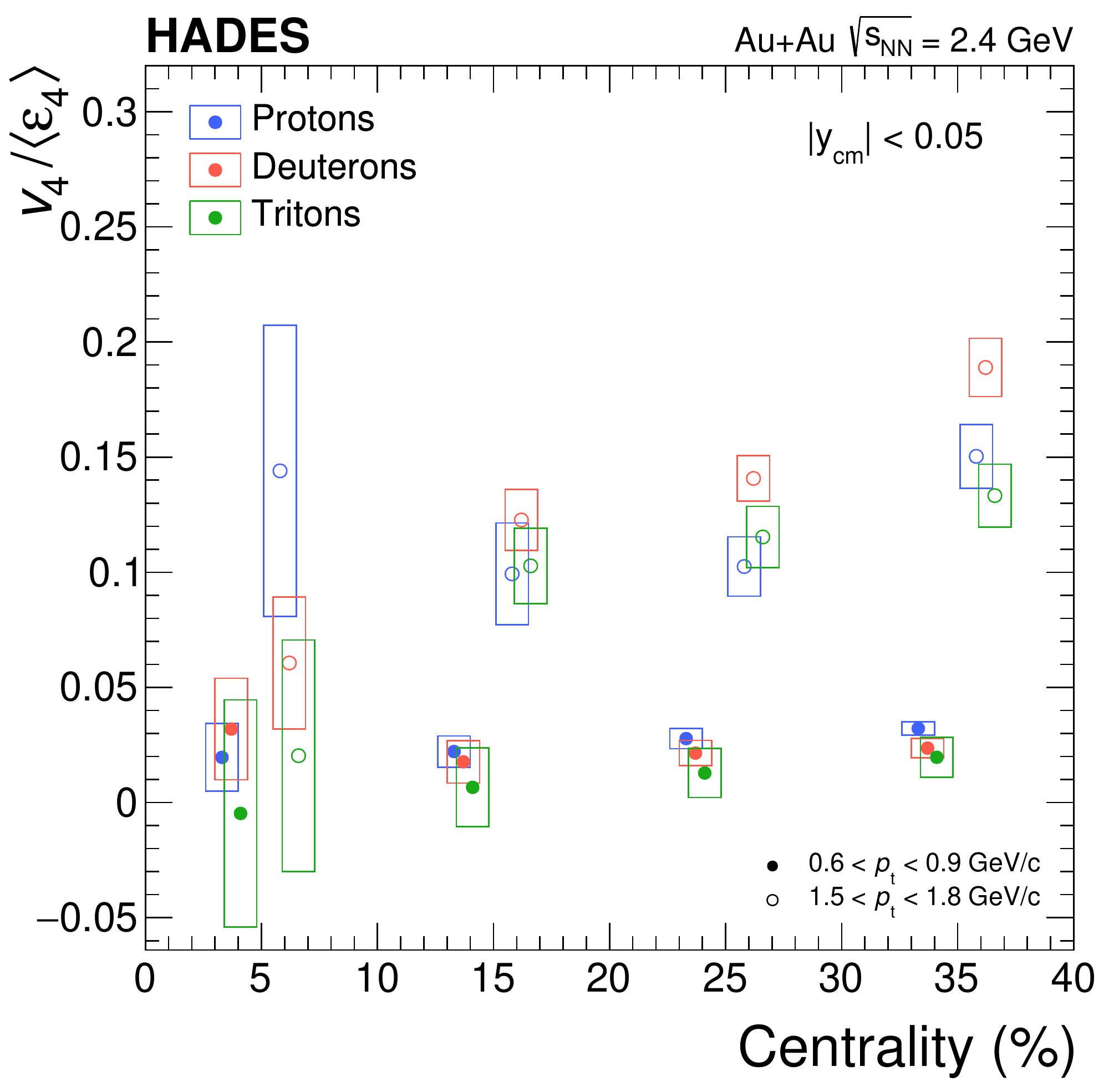}
\end{center}
\caption{Same as in \Fi{fig:v2_eccScale}, but for the scaled 
   quadrangular flow. The values are divided by the square of second
   order eccentricity, $\vfour / \etwoav^{2}$ (left panel), and by the
   fourth order eccentricity, $\vfour / \efourav$ (right panel).}
\label{fig:v4_eccScale}
\end{figure*}
%

%
\begin{figure*}
\begin{center}
\includegraphics[width=0.40\textwidth]{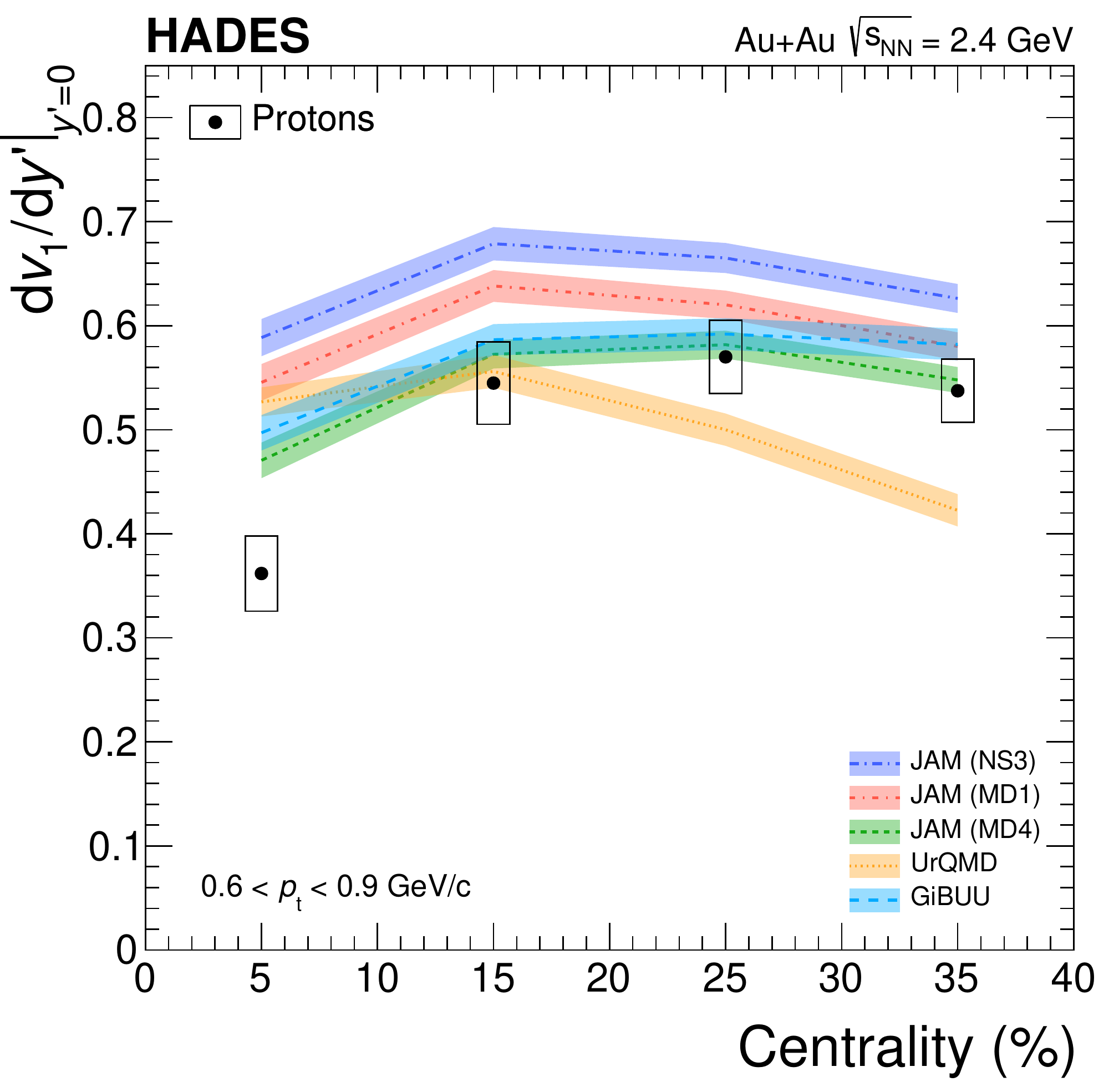}
\includegraphics[width=0.40\textwidth]{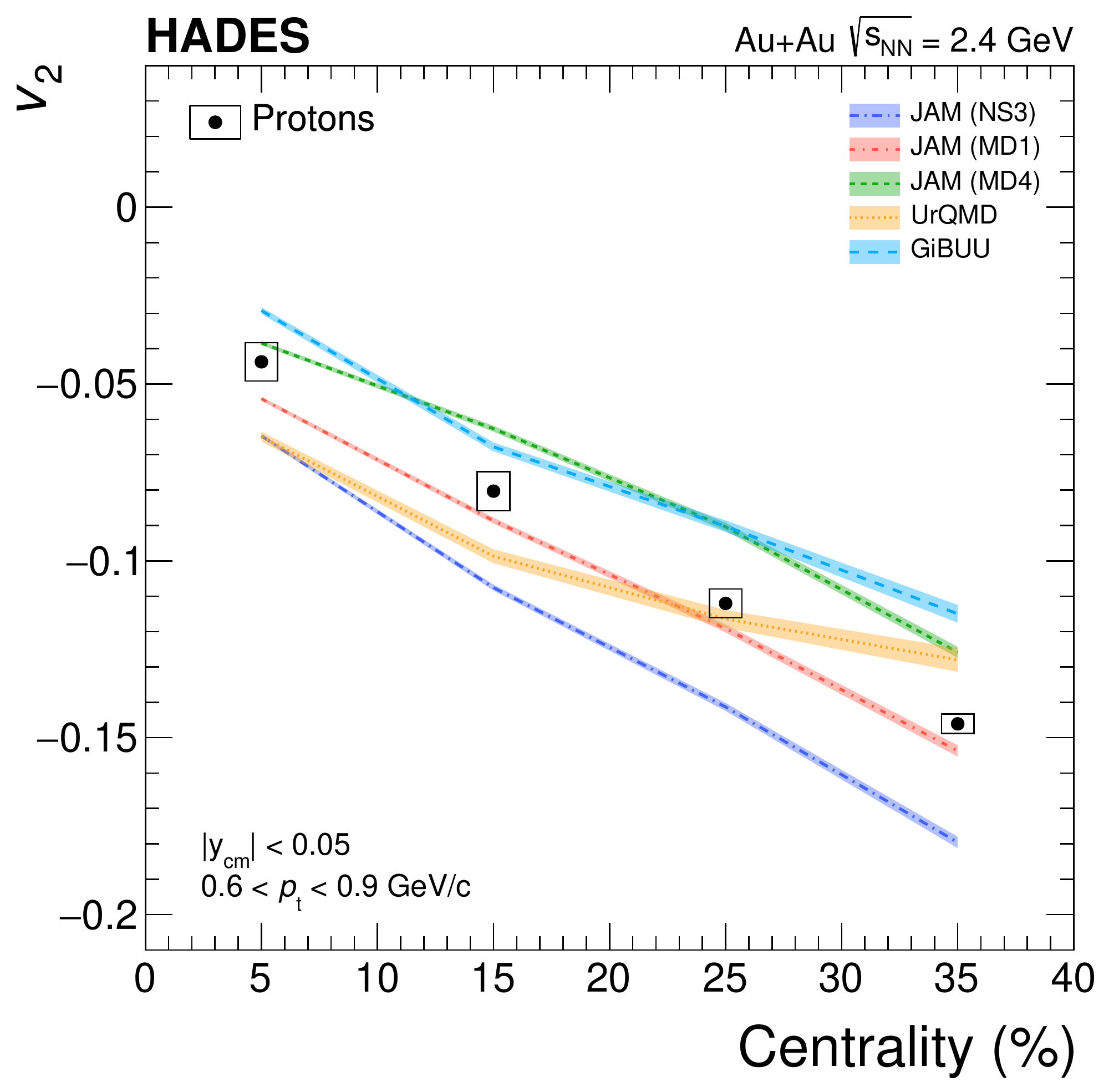}
\includegraphics[width=0.40\textwidth]{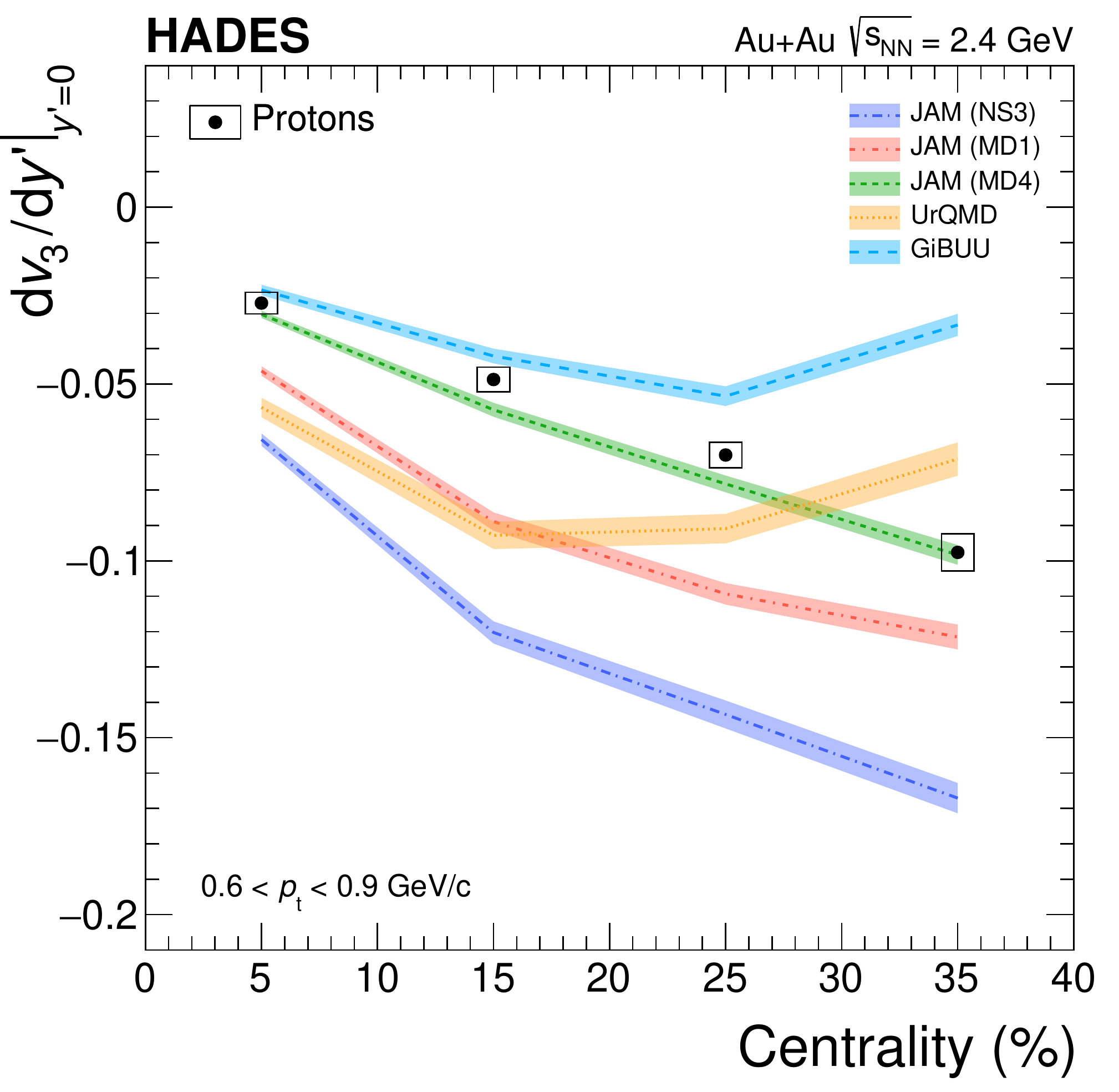}
\includegraphics[width=0.40\textwidth]{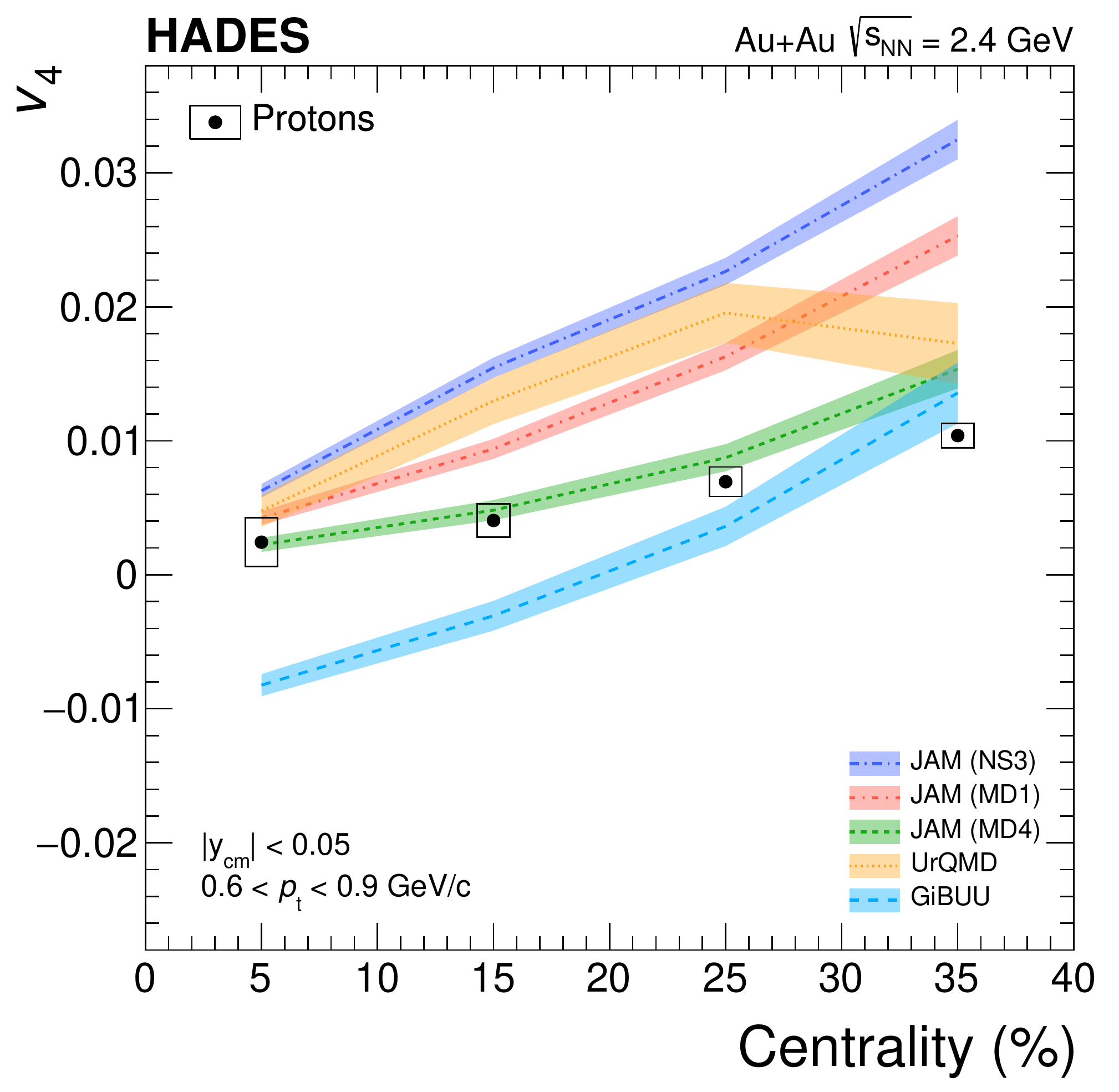}
\end{center}
\caption{Directed \dvonemid\ (left top panel), elliptic \vtwo\
  (right top panel), triangular \dvthrmid\ (left bottom panel) and
  quadrangular \vfour\ (right bottom panel) flow of protons in the
  transverse momentum interval $0.6 < \pt < 0.9$~\gevc\ at
  mid-rapidity in Au+Au collisions at $\sqrtsnn = 2.4$~GeV for four
  centrality classes.  The data are compared to several model
  predictions (see text for details). The width of the bands reflect
  the statistical uncertainties of the model calculations.}
\label{fig:vn_p_lowpt_model}
\end{figure*}
%

%
\begin{figure*}
\begin{center}
\includegraphics[width=0.40\textwidth]{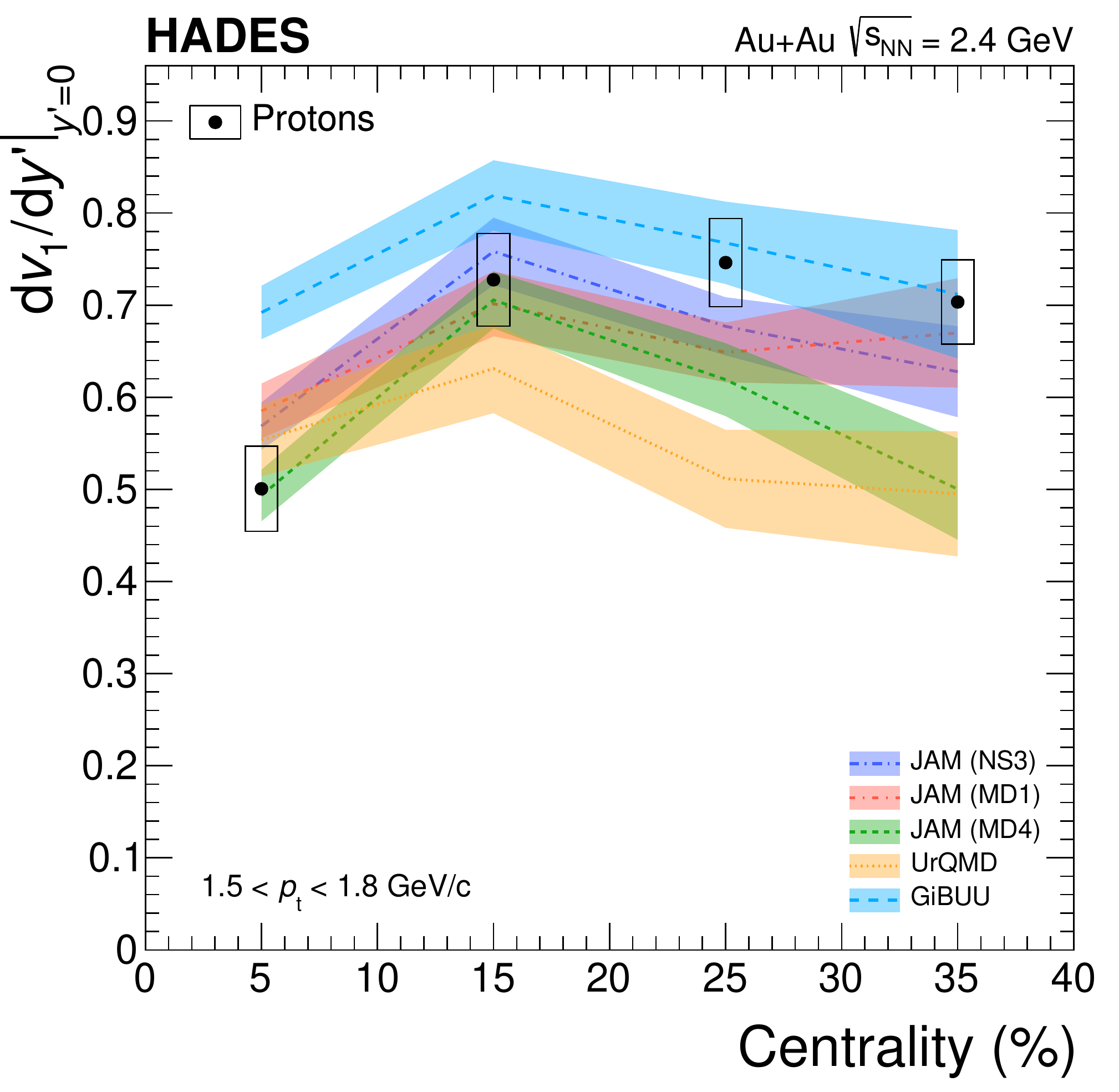}
\includegraphics[width=0.40\textwidth]{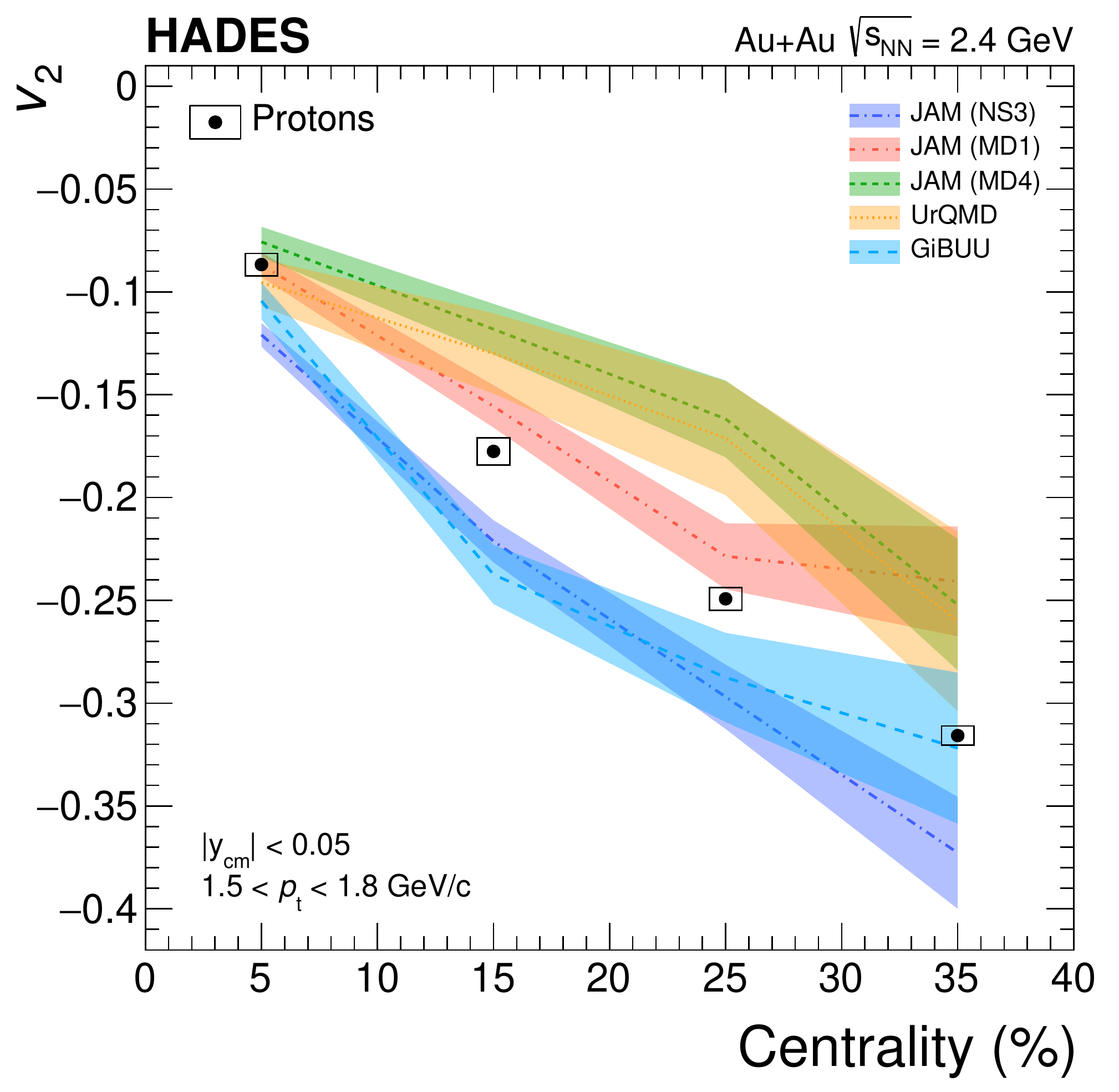}
\includegraphics[width=0.40\textwidth]{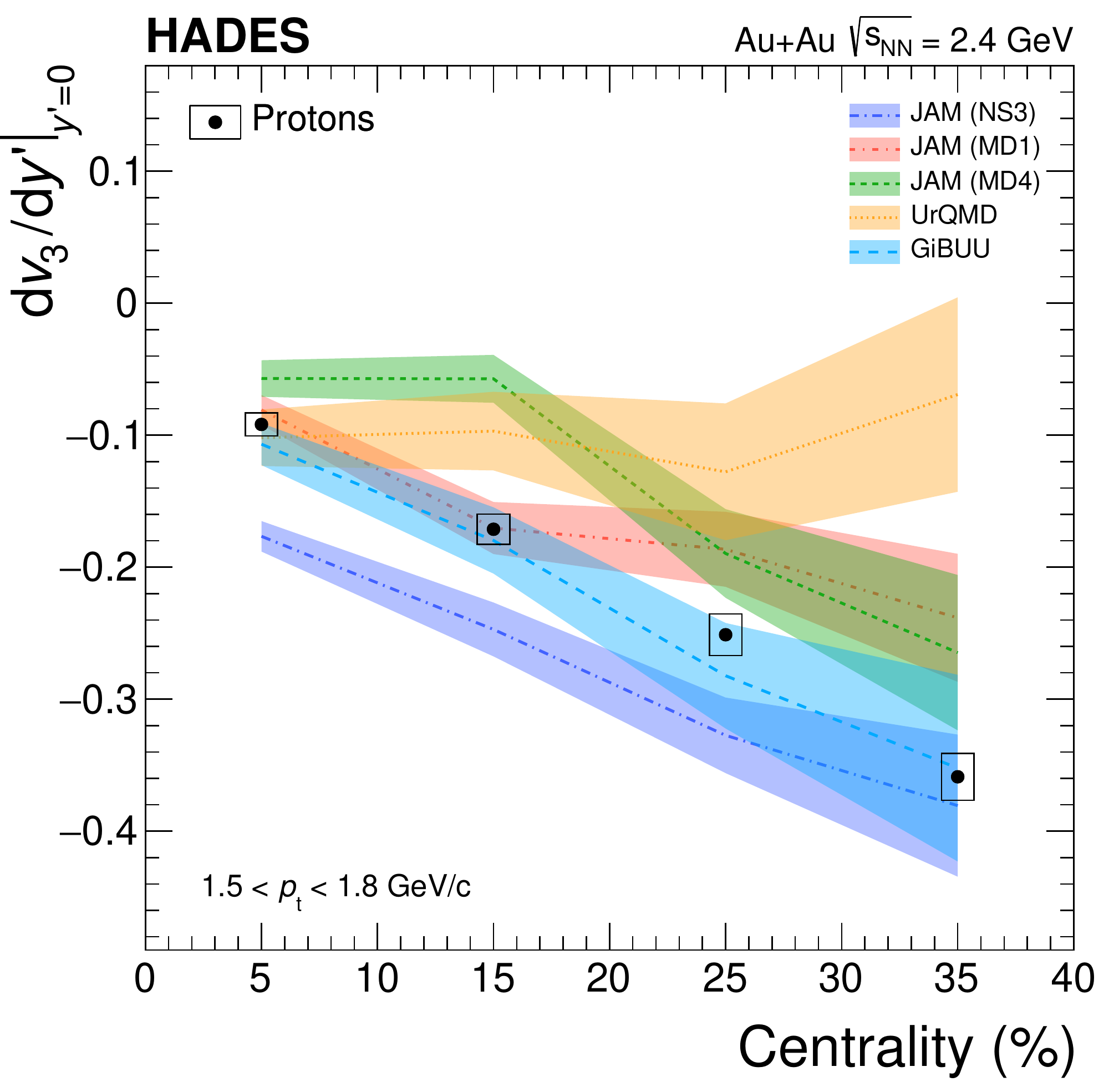}
\includegraphics[width=0.40\textwidth]{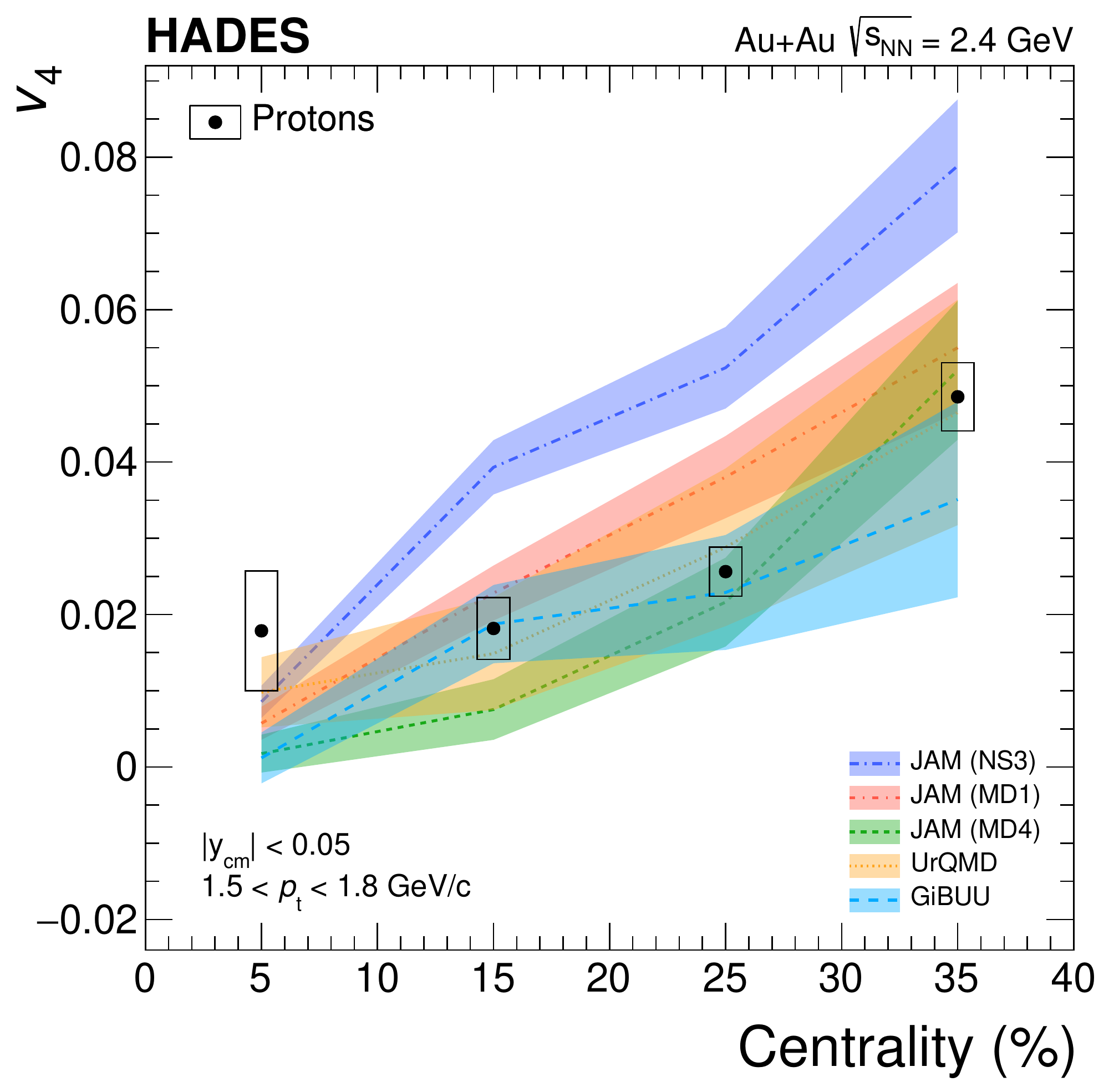}
\end{center}
\caption{Same as in \Fi{fig:vn_p_lowpt_model}, but for the transverse
  momentum interval $1.5 < \pt < 1.8$~\gevc.}
\label{fig:vn_p_highpt_model}
\end{figure*}
%

%
\section{Scaling properties}
\label{sect:scaling}

\subsection{Coalescence scaling}
\label{sect:coal}

A comparison of the \pt~dependences of \vtwo\ measured at mid-rapidity
for protons, deuterons and tritons (see \Fi{fig:v2_all_scaling_20-30})
demonstrates a clear mass ordering $\vtwo^{\rb{prot.}} <
\vtwo^{\rb{deut.}} < \vtwo^{\rb{trit.}}$ for $\pt < 1.5$~\gevc.
Within a naive nucleon-coalescence scenario one would expect that the
observed flow coefficients scale with the nuclear mass number $A$
according to the relation $v_{n,A}(A \, \pt) =$ \linebreak $ A \,
\vn(\pt)$, where \pt\ is the momentum of a single nucleon and $p_{t,A}
= A \pt$ the one of the composite nuclei.  The correspondingly scaled
\pt~dependences of the proton \vtwo\ are shown in
\Fi{fig:v2_all_scaling_20-30} as solid curves for $A = 2$ and $3$.
The agreement with the measured \vtwo\ values for deuterons and
tritons is already quite good.  However, this approximation only holds
for small flow values and, as \vtwo\ measured at high \pt\ is quite
sizeable, a correction term has to be taken into account
\cite{Molnar:2003ff, Kolb:2004gi}:
\begin{eqnarray}
 v_{n,A=2}(2 \, \pt) & = & 2 \, \vn(\pt) \;
                          \frac{1}{1 + 2 \, \vn^{2}(\pt)} \, ,
\nonumber \\
 v_{n,A=3}(3 \, \pt) & = & 3 \, \vn(\pt) \;
                          \frac{1 + \vn^{2}(\pt)}{1 + 6 \, \vn^{2}(\pt)}
\, .
\label{eq:vn_coal_full}
\end{eqnarray}
In fact, the correspondingly scaled proton \vtwo~values agree well
with the ones measured for deuterons and tritons up to the highest
\pt, as depicted by the coloured bands in
\Fi{fig:v2_all_scaling_20-30}.  It should be noted though, that this
kind of scaling is only observed in the region around mid-rapidity,
where the elliptic flow is the predominant component of the azimuthal
distributions. 

Similar to the case of \vtwo\ (see \Fi{fig:v2_all_scaling_20-30}) also
for \vfour\ a nuclear mass scaling behaviour is observed at
mid-rapidity.  This is demonstrated in \Fi{fig:v4_scaling_all_20-30},
which shows a comparison of \vfour\ at mid-rapidity as a function of
\pt\ for protons, deuterons and tritons.  A mass hierarchy can be
observed, $\vfour^{\rb{prot.}} > \vfour^{\rb{deut.}} >
\vfour^{\rb{trit.}}$, at least in the lower \pt~region.  In order to
test whether this ordering of \vfour\ is also compatible with a
nucleon coalescence scenario, we use an extension of
\Eq{eq:vn_coal_full} which takes combinations of different orders into
account\footnote{Please note that here mixed terms between \vtwo\
  and \vfour\ are ignored.}:
\begin{eqnarray}
 v_{4,A=2}(2 \, \pt) & = & \frac{2 \, \vfour(\pt) + \vtwo^{2}(\pt)}
                               {1 + 2 \, \vtwo^{2}(\pt)
                                  + 2 \, \vfour^{2}(\pt)}
 \, ,
 \nonumber \\
 v_{4,A=3}(3 \, \pt) & = & \frac{3 \, \vfour(\pt) + 3 \, \vtwo^{2}(\pt)}
                               {1 + 6 \, \vtwo^{2}(\pt) 
                                  + 6 \, \vfour^{2}(\pt)}
\, .
\label{eq:v4_coal_with_v2}
\end{eqnarray}
Assuming the relation $\vtwo = - \sqrt{2 \, \vfour}$~\cite{Adamczewski-Musch:2020iio} this reduces to
\begin{eqnarray}
 v_{4,A=2}(2 \, \pt) & = & 4 \, \vfour(\pt) \: 
                          \frac{1}
                               {1 + 4 \, \vfour(\pt)
                                  + 2 \,  \vfour^{2}(\pt)}
\, ,                           
\nonumber \\
 v_{4,A=3}(3 \, \pt) & = & 9 \, \vfour(\pt) \:
                          \frac{1}
                               {1 + 12 \, \vfour(\pt)
                                  + 6 \, \vfour^{2}(\pt)}
\, .
\label{eq:v4_full_withv4v22}
\end{eqnarray}
If the higher-order correction is omitted, this results in the simple
approximation of $v_{4,A}(A \, \pt) = A^{2} \, \vfour(\pt)$, which
therefore should only be valid for small flow values.
Figure~\ref{fig:v4_scaling_all_20-30} includes a comparison of these
different approximations to the data.  While the relation given in
\Eq{eq:vn_coal_full} does not provide a good match with the data,
its extended version given in \Eq{eq:v4_full_withv4v22} results in a
very good description of the deuteron and triton data.  Also, the
simple relation $v_{4,A}(A \, \pt) = A^{2} \, \vfour(\pt)$ is quite
close to the data points, indicating that the higher-order corrections
are small.

While the above discussed scaling properties can be indicative for
nucleon coalescence as the main process responsible for light nuclei
formation, a more refined discussion would involve the comparison to
various models.  Examples for implementations of the coalescence
approach within transport models to describe HADES and STAR flow data,
respectively, can be found in \cite{Liu:2019nii,Hillmann:2019wlt}.
These studies should be extended in the future in a more systematical
way using several transport models in order to arrive on firmer
conclusions on this topic.

\subsection{Initial eccentricity}
\label{sect:ecc}

%
\begin{table}
\footnotesize
\begin{center}
\begin{tabular}{lccc}
\hline
Centrality          &
$\bav$              &
$\etwoav$           &
$\efourav$          \\
\hline
$00-10$ & $3.13$ & $0.121 \pm 0.007$ & $0.124 \pm 0.009$ \\
$10-20$ & $5.70$ & $0.235 \pm 0.010$ & $0.183 \pm 0.009$ \\
$20-30$ & $7.37$ & $0.325 \pm 0.008$ & $0.250 \pm 0.010$ \\
$30-40$ & $8.71$ & $0.401 \pm 0.009$ & $0.323 \pm 0.012$ \\
\hline
\end{tabular}
\end{center}
\caption{Parameters describing the initial nucleon distribution for
  the different centrality classes as calculated within the Glauber-MC
  approach~\cite{Adamczewski-Musch:2017sdk}.  Listed are the
  corresponding average impact parameter $\langle b \rangle$ and
  the average participant eccentricities \etwoav\ and \efourav.}
\label{tab:ecc2GlauberMC}
\end{table}

In order to investigate to what extent the spatial distribution of the
nucleons in the initial state of the collision system determines the
observed flow pattern, we use the eccentricity $\epsilon_{n}$ of order
$n$ of the participant nucleon distribution in the transverse plane as
calculated within the Glauber-MC
approach~\cite{Adamczewski-Musch:2017sdk,Loizides:2017ack}:
\begin{equation}
   \epsilon_{n} = \frac{\sqrt{\langle r^{n} \cos(n\phi) \rangle^{2}
                            + \langle r^{n} \sin(n\phi) \rangle^{2}}}
                       {\langle r^{n} \rangle}
  \, ,
\end{equation}
with $r = \sqrt{x^{2} + y^{2}}$, $\phi = \tan^{-1} (y / x)$ and
$x$, $y$ as the nucleon coordinates in the plane perpendicular to the
beam axis, where $x$ is oriented in the direction of the impact
parameter.  The values calculated for the different centrality classes
are given in \Ta{tab:ecc2GlauberMC}.

Figure~\ref{fig:v2_eccScale} shows the elliptic flow measured at
mid-rapidity for all three investigated particle species after
dividing it by the event-by-event averaged second-order participant
eccentricity, $\vtwo/\etwoav$.  Remarkably, this scaling results in
almost identical values for all centrality classes at high transverse
momenta, indicating that the centrality dependence of the elliptic
flow of particles emitted at early times is to a large extent already
determined by the initial nucleon distribution.  However, as the
elliptic flow at these beam energies is due to the so-called
squeeze-out effect, caused by the passing spectators, it is not
immediately clear how the flow pattern can be directly related to the
initial participant distribution.  A possible explanation might be
that the distribution of the spectators forms a negative image of the
one of the participants and thus could imprint its shape onto the
emission pattern of the light nuclei.  The scaling of \vtwo\ works
less well at lower \pt, which suggests that particles emitted at later
times are less affected by the initial state geometry.

Also, we observe a scaling of \vfour\ with $\etwo^{2}$, as depicted
in the left panel of \Fi{fig:v4_eccScale} which presents $\vfour
/ \etwoav^{2}$ for different centralities in two transverse momentum
intervals.  This points to a fixed relation between \vtwo\ and \vfour,
such that the latter is a second order correction $\propto \etwo^{2}$
to the overall emission pattern defined at mid-rapidity by \vtwo.
This is contrary to the case at very high collision energies, where
higher-order flow coefficients are related to initial state
fluctuations and thus, to a large extent, should be independent of one
another.  In this scenario one would also expect \vfour\ to scale
rather with \efour.  While this might be observable also here at lower
\pt, $\vfour / \efourav$ is not independent of centrality in the high
\pt~region, i.e. for particles emitted at early times, as demonstrated
in the right panel of \Fi{fig:v4_eccScale}.
%

%
\section{Model comparisons}
\label{sect:models}

In the following, several transport model calculations are compared
with the measured flow data.  These models provide the possibility to
test the effect of the EOS of dense nuclear matter on the flow
coefficients by implementing different density dependent potentials.
Usually, these are parameterised such that the dependence on the
baryon density $\rho$ results in either a weak (``soft EOS'') or a
strong (``hard EOS'') repulsion of compressed nuclear matter.  The
comparison with data then allows for a discrimination between these
two scenarios.  While previous investigations were only based on
measurements of the directed and elliptic flow
\cite{Shi:2001fm,LeFevre:2016vpp}, the information from higher-order
flow coefficients will provide additional discriminating power.
Ultimately, the multi-differential high-statistics data presented here
\linebreak
should enable a direct extraction of the EOS parameters via a Bayesian
fit of the models to the data.  However, as a prerequisite it is
important to establish that the various model approaches do not differ
significantly in their predictions in order to allow for a consistent
determination of the EOS. For a detailed review of the different
approaches used for transport simulations see~\cite{TMEP:2022xjg,
Colonna:2021xuh, Xu:2016lue}.  As examples the predictions by two QMD
models, JAM~1.9~\cite{Nara:2020ztb} and
UrQMD~3.4~\cite{Hillmann:2018nmd}, and one BUU model,
GiBUU~2019~\cite{Buss:2011mx} are considered here. The JAM code is used
with three different EOS implementations: hard momentum independent
NS3, hard momentum dependent MD1 and soft momentum dependent MD4. The
UrQMD code is employed with the ``hard EOS'', and GiBUU with the
``soft EOS'' (Skyrme~12).

Comparisons of the model predictions to the proton flow of different
order measured at mid-rapidity are presented, as a function of
centrality, in \Fi{fig:vn_p_lowpt_model} (low \pt~interval $0.6 < \pt <
0.9$~\gevc) and \Fi{fig:vn_p_highpt_model} (high \pt~interval $1.5 <
\pt < 1.8$~\gevc). As most models do not include a dedicated mechanism
for the generation of light clusters, which would be needed for a
realistic prediction of deuteron and triton flow, we restrict the
comparison here to protons only.

Generally, all models roughly capture the overall magnitude and trend
of the measured data. In the lower \pt~region the differences between
the models are relatively small. JAM with MD4 provides the overall best
reproduction of the data points, with the exception of \vtwo\ where MD1
is closer to the data. UrQMD is close to the data for \vtwo, but
deviates for \vone, \vthree\ and \vfour\ at several centralities, while
GiBUU reproduces generally better with the exception of \vfour. In the
higher \pt~interval the deviations between the models are a bit larger.
Here JAM with MD1 yields the best match to the data, while MD4 and NS3
do not provide a consistent description of the measurements. Also, for
UrQMD and GiBUU, systematic deviations are observed for some orders of
the flow coefficients. Nevertheless, the models presented here should
form a useful basis for further, more detailed data comparisons and and
consistent determination of the EOS. It should be noted that not all
model calculations include the effects of momentum and isospin
dependent potentials, which, however, will be essential for this
purpose. Furthermore, a common treatment of cluster formation should be
implemented which will allow for an usage of the data on deuteron and
triton flow as an additional constraint.

%

%
\section{Conclusions}
\label{sect:conclusions}

In summary, we have presented a detailed multi-diffe\-ren\-tial
measurement of collective flow coefficients of protons, deuterons 
and tritons in Au+Au collisions at $\sqrtsnn = 2.4$~GeV using the 
high-statistics data set collected with the HADES experiment.  The
directed (\vone), elliptic (\vtwo) and higher order (\vthree\ and
\vfour) flow coefficients were determined with respect to the
first-order event-plane measured at projectile rapidities.  The
centre-of-mass energy of $\sqrtsnn = 2.4$~GeV is close to the region
where $\dvonedyp|_{y^{\prime} = 0}$ is maximal and $\vtwo$ is minimal.
All flow coefficients were extracted as a function of transverse
momentum \pt\ and rapidity \ycm\ over a large region of phase-space
and in four centrality classes.  The \pt\ and \ycm\ dependences of
\vone\ are very similar in shape for protons, deuterons and tritons.
A clear mass hierarchy is observed for \vone\ values measured away
from mid-rapidity at higher \pt, as well as for
$\dvonedyp|_{y^{\prime} = 0}$, which both increase with the mass of
the particle.  The elliptic flow coefficient \vtwo\ has a Gaussian
shaped rapidity distribution, whose width narrows with increasing
particle mass, such that the rapidity value for which \vtwo\ changes
sign moves closer towards mid-rapidity for increasing mass number.
Both, the proton directed flow $\dvonedyp|_{y^{\prime} = 0}$ and the
elliptic flow \vtwo\ at mid-rapidity are in line with the established
energy dependence.  The \pt\ and \ycm\ dependences of \vthree\
(\vfour) are relatively similar to the one of \vone\ (\vtwo), but have
the opposite sign.  At mid-rapidity a nucleon number scaling is
observed in the \pt~dependence of \vtwo\ (\vfour), when dividing the
values of \vtwo\ (\vfour) by $A$ ($A^{2}$) and \pt\ by $A$.  This
might be indicative for nuclear coalescence as the main process
responsible for light nuclei formation.  Such a straightforward
scaling is not seen at more forward and backward rapidities.  The
elliptic flow measured at mid-rapidity at higher \pt\ is found to be
independent of centrality for all three investigated particle species
after dividing it by the event-by-event averaged second order
participant eccentricity $\vtwo/\etwoav$.  A similar scaling is
observed for \vfour\ after division by $\etwo^{2}$.

The new multi-differential high-precision data on \vone, \vtwo,
\vthree, and \vfour\ provides important constraints on the
equation-of-state of compressed baryonic matter as used in models of
relativistic nuclear collisions \cite{Huth:2021bsp}.  In particular,
the higher moments provide more discriminating power than the directed
and elliptic flow alone.  The general features of the data on proton
flow at mid-rapidity are qualitatively captured by several transport
models.  A consistent and exact description of all flow coefficients
over the whole phase-space and at all investigated centralities is not
yet possible.  With further progress in the theoretical developments
it should be feasible to use the data shown here, together with other
measurements, to directly extract a precise parametrization of the
equation-of-state of compressed nuclear matter.

%

%
\begin{acknowledgements}
  The collaboration gratefully acknowledges the support by
  SIP JUC Cracow, Cracow (Poland), 2017/26/M/ST2/00600; TU Darmstadt,
  Darmstadt (Germany), VH-NG-823, DFG GRK 2128, DFG CRC-TR 211,
  BMBF:\linebreak 05P18RDFC1; Goethe-University, Frankfurt (Germany)
  and TU Darmstadt, Darmstadt (Germany), DFG GRK 2128, DFG CRC-TR 211,
  BMBF:05P18RDFC1, HFHF, ELEMENTS:\linebreak 500/10.006, VH-NG-823,
  GSI F\&E, ExtreMe Matter Institute EMMI at GSI Darmstadt; JLU
  Giessen, Giessen (Germany), BMBF:05P12RGGHM; IJCLab Orsay, Orsay
  (France), CNRS/IN2P3; NPI CAS, Rez, Rez (Czech Republic), MSMT
  LTT17003, MSMT LM2018112, MSMT OP VVV
  CZ.02.1.01/\linebreak 0.0/0.0/18\_046/0016066. \\
  \\
  The following colleagues from Russian institutes did contribute to
  the results presented in this publication but are not listed as
  authors following the decision of the HADES collaboration Board on
  March 23: A.~Belyaev, S.~Chernenko, O.~Fateev, O.~Golosov,
  M.~Golubeva, F.~Guber, A.~Ierusalimov, A.~Ivashkin, A.~Kurepin,
  A.~Kurilkin, P.~Kurilkin, V.~Ladygin, A.~Lebedev, M.~Mamaev,
  S.~Morozov, O.~Petukhov, A.~Reshetin, A.~Sadovsky.
\end{acknowledgements}
%

%
\bibliographystyle{apsrev4-2}     
\bibliography{hades_flow}
%
%
\end{document}